\documentclass{statsoc}\usepackage[]{graphicx}\usepackage[]{color}
\makeatletter
\def\maxwidth{ %
  \ifdim\Gin@nat@width>\linewidth
    \linewidth
  \else
    \Gin@nat@width
  \fi
}
\makeatother

\definecolor{fgcolor}{rgb}{0.345, 0.345, 0.345}

\usepackage{framed}
\makeatletter
 {\par\unskip\endMakeFramed%
 \at@end@of@kframe}
\makeatother

\definecolor{shadecolor}{rgb}{.97, .97, .97}
\definecolor{messagecolor}{rgb}{0, 0, 0}
\definecolor{warningcolor}{rgb}{1, 0, 1}
\definecolor{errorcolor}{rgb}{1, 0, 0}
\newenvironment{knitrout}{}{} % an empty environment to be redefined in TeX

\usepackage{alltt}

\usepackage{url}
\usepackage{xcolor}

\usepackage{amssymb}

\usepackage{bm}
\usepackage{amsmath,array}
\usepackage[a4paper%left=2cm, top=2cm, bottom=2cm, right=2cm
]{geometry}
\usepackage{natbib}
\usepackage{graphicx}

\usepackage{soul}
\usepackage{enumerate}
\usepackage{float}
\restylefloat{table}
\definecolor{Set1zwei}{RGB}{55, 126, 200}%vgl. Set1 (RColorBrewer)
\definecolor{Set1drei}{RGB}{77, 190, 90}
\definecolor{Set2drei}{RGB}{141, 160, 203}%vgl. Set2 (RColorBrewer)
\definecolor{Set2fuenf}{RGB}{205, 255, 200}
            
\definecolor{myc}{RGB}{0,0,0} %57, 125, 243}

\usepackage[
    colorlinks,
    linkcolor={red!50!black},
    citecolor={blue!50!black},
    urlcolor={blue!80!black}
    ]{hyperref}

\usepackage{float}
\floatstyle{plaintop}
\restylefloat{table}
\usepackage[tableposition=top,skip=2pt]{caption}
\usepackage{setspace}
\title[Boosting Factor-Specific Historical Models]{Boosting Factor-Specific Functional Historical Models for the Detection of Synchronisation in Bioelectrical Signals}

\author[R\"ugamer {\it et al.}]{David R\"ugamer}
\address{Department of Statistics, LMU Munich,
Munich,
Germany.}
\author{Sarah Brockhaus}
\address{Department of Statistics, LMU Munich,
Munich,
Germany.}
\author{Kornelia Gentsch}
\address{Swiss Center for Affective Sciences, University of Geneva,
Geneva,
Switzerland.}
\author{Klaus Scherer}
\address{Swiss Center for Affective Sciences, University of Geneva,
Geneva,
Switzerland.}
\author[R\"ugamer {\it et al.}]{Sonja Greven}
\address{Department of Statistics, LMU Munich,
Munich,
Germany.}
\IfFileExists{upquote.sty}{\usepackage{upquote}}{}
\begin{document}
\vspace*{-0.15cm}
% \begin{footnotesize}

%\vspace*{-0.15cm}
% A short summary (no more than about 100 words) should be included at the beginning of the manuscript, together with five or six keywords or key phrases,
\begin{abstract}
The link between different psychophysiological measures during emotion episodes is not well understood. To analyse the functional relationship between electroencephalography (EEG) and facial electromyography (EMG), we apply historical function-on-function regression models to EEG and EMG data that were simultaneously recorded from 24 participants while they were playing a computerised gambling task. Given the complexity of the data structure for this application, we extend simple functional historical models to models including random historical effects, factor-specific historical effects, and factor-specific random historical effects. Estimation is conducted by a component-wise gradient boosting algorithm, which scales well to large data sets and complex models.
\end{abstract}
% \end{footnotesize}


% read all objects

\newpage

\section{Introduction}

%Bioelectrical signals or biosignals have become more and more important in recent years. 
Bioelectrical signals such as electromyography (EMG), electroencephalography (EEG) or electrocardiogram (ECG) are variations in electrical energy that carry information about living systems \citep{Semmlow.2014}. An appropriate analysis of bioelectrical signals, usually obtained in the form of time series data, is a crucial point in many research areas, including (tele-)me\-di\-cine, automotive technology, and psychology \citep{Kang.2006, Kaniusas.2012}.
%'  
%' <<pubmed, echo=FALSE, message=FALSE, comment=NA, fig.height=3.5, fig.width=6.5, warning=FALSE, fig.cap="Number of articles associated with the phrase 'biosignal' in the pubmed data base (\\url{http://www.ncbi.nlm.nih.gov/pubmed}, as of 21 June 2016).", fig.scap = NA, fig.pos='ht', fig.align='center'>>=
%' dat <- read.csv2("timeline.csv", sep = ",", header = T)
%' library(ggplot2)
%' ggplot(dat, aes(x=year, y=count)) + geom_bar(stat="identity") + theme_bw() + theme(text = element_text(size=12)) + scale_x_continuous(breaks=c(1980,1990,2000,2010,2016),labels=c(1980,1990,2000,2010,2016))
%' @
In the field of cognitive affective neuroscience, a particular interest lies in the link of measured brain activity recorded with the EEG, and peripheral response systems such as the heart rate or facial muscle activity. %synchronisation analysis of time series has become important as 
%technological development results in an increasing amount of available data. 
%The dynamically changing electrical activity, measured at several locations on the surface of the sculp with the EEG, for example, may yield time series realisations with a high sampling rate and %thus, many data points. A 
%a current challenge in the analysis of such data includes the detection of phases, in which multiple signals synchronise \citep[see e.g.][Chap. 25]{Cohen.2014}.
In this context, our motivating study \citep{Gentsch.2014} investigated the coherence between emotion components. In componential emotion theory, an emotional episode is thought to be an emergence of coherent or temporally correlated changes in emotion components, such as appraisals or facial expressions. This is referred to as synchronisation \citep{Grandjean.2009}.

\textit{The emotion components data}. In the study of \citeauthor{Gentsch.2014}, brain activity (EEG) as well as facial muscle activity (EMG) was simultaneously recorded. The data set at hand consists of time series of $384$ equidistant observed time points for both EEG and EMG signals, eight different study settings (conditions in a computerised gambling game) and $24$ participants. %For each participant and each setting, more than $100$ measured time series of the $64$ EEG electrodes and the EMG of three facial muscles are available, yielding a total number of $20,718$ time series for each EEG and EMG electrode. After pre-processing (see Section \ref{application} for more details), every time series is given by $384$ equidistant observed time points within $1700ms$ ($256Hz$). 
The traditional approach of analysing EEG and EMG data is to calculate the average signal for each participant across all trials of one study setting. For EEG data, this is referred to as event-related potential analysis \citep[see, e.g.,][]{Pfurtscheller.1999}. Such an aggregation yields a reduced data set of $N = 8 \cdot 24 \cdot 384 = 73,728$ observed data points. At each of the $N$ time points, measurements are available for three EMG and $64$ EEG electrodes. 
Figure \ref{fig:explData} depicts one EEG and EMG signal for one participant and all eight study settings with a common starting point of $200ms$ after stimulus onset.%, the EEG signals of different settings yield similarities in particular between $150$ and $300ms$. %During this period the participant is evaluating the given situation, resulting in more distinctive signals afterwards. \color{red}Passt das so? \color{black} 
%For the EMG signal, differences between the signals of different game scenarios are more subtle. 

Efferent signals from the brain \textcolor{myc}{(signals originating from the brain)} innervate \textcolor{myc}{or activate} facial muscles \citep[see, e.g.,][]{Rinn.1984}. Therefore, it should be possible to trace back facial muscle activity recorded with facial EMG to brain activity captured with EEG. As certain cognitive processes can be related to different brain areas and facial regions, our particular interest lies in investigating the link between a selected EEG electrode signal and a specific EMG signal. We expect any association between these two signals to (a) be time-varying, (b) exhibit a temporal lag that is a priori unknown (even though a minimum lag can be inferred from the literature), (c) be specific to a study setting and / or (d) be only present during certain time intervals. 
 %In order to verify our results, we also use a more coarse aggregation, yielding a multiple of $N$ observations.  
\begin{knitrout}
\definecolor{shadecolor}{rgb}{0.969, 0.969, 0.969}\color{fgcolor}\begin{figure}[ht]

{\centering \includegraphics[width=\maxwidth]{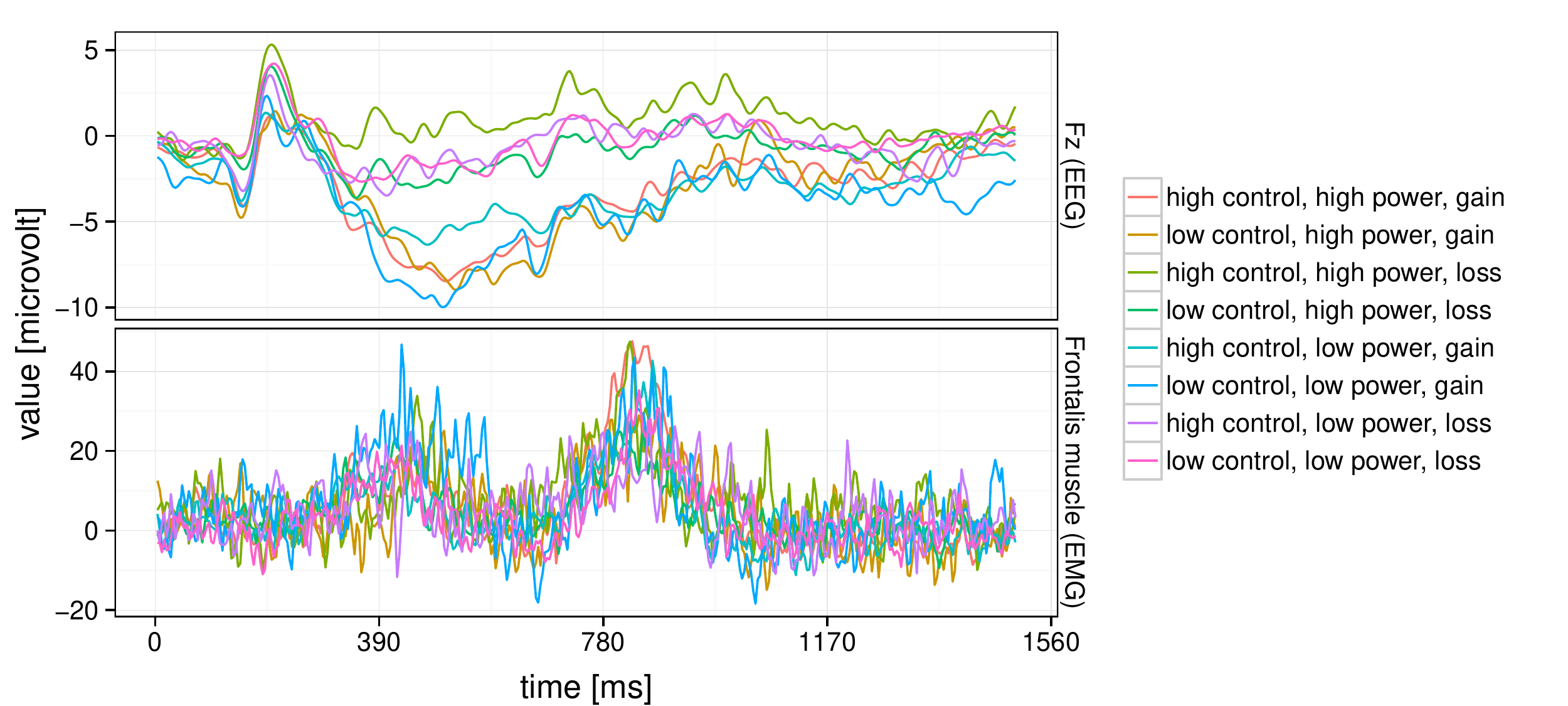} 

}

\caption{Example for one EEG signal (Fz electrode) and one EMG signal (frontalis muscle for raising eyebrows) of one participant, averaged over all trials for each of the eight possible game conditions (line colours)}\label{fig:explData}
\end{figure}

\end{knitrout}

\textit{Existing methods for detecting synchronisation}. 
%Analogously to the standard correlation coefficient, measures of association between two time series can, for example, be defined via the cross-correlation function or the coherence. Whereas the cross-correlation is a function of the time lag $m = 0, \pm 1, \pm 2, \ldots$ between the two series in time-domain analysis, the coherence is a function of the frequency in frequency-domain analysis \citep[see e.g.][]{Pawitan.2005}.
Previous approaches to detect synchrony in brain activity and autonomic physiology data have mostly focused on coherence or cross-cor\-re\-la\-tion. Examinations of EEG and EMG synchronisation can, inter alia, be found in \citet{biolIssue, Mima.1999, Brown.2000, Mima.2000a, Mima.2000b, Grosse.2002, Quiroga.2002, Bortel.2006, Hashimoto.2010}. While coherence is a function of the frequency measuring the explained variance of one time series by another time series in the frequency-domain, cross-correlation is a function of time, yielding the correlation of two time series for a given lag \citep[see, e.g.,][]{Pawitan.2005}.  %Established methods for the analysis of different EEG signals are presented in \citet{Cohen.2014}. A different approach is, for example, given by \citet{Quiroga.2002}, proposing a simple and fast method to measure synchronisation between signals by combining counts of events. 
%Though it is possible to perform frequency-domain analysis on predefined time sections of each signal and hence to infer on time-dependent synchronisation of two signals in these sections, 
With the aim to relate different time points of two signals to each other, we focus on methods in the time-domain. Established methods are, however, concerned with the estimation of the association between two observed time series rather than the analysis of a large number of time series observations given in pairs of two signals. This applies for (cross-)correlation, which additionally does not provide the possibility to take covariates into account, as well as for other methods such as the \emph{generalised synchronisation} approach based on the state-space representation \citep{Diab.2013} or autoregressive times series approaches \citep[see e.g.][]{Ozaki.2012}. Furthermore, most of these approaches require the definition of a specific or a maximum time lag.  %Hence, such measures are limited by their nature relating one time point (or frequency) in one series to only one time point (respectively frequency) of the second series. 
% In addition, correlation approaches do not provide the possibility to take covariates into account.

\textit{Function-on-function regression}. 
%As both the EEG and the EMG signal can be seen as realisations of an intrinsically smooth process, another possibility to describe and infer the relationship of such ``time series'' is given by functional data analysis. The independent as well as the dependent variable consist of temporally dependent observations and can be understood as noisy observations of functional variables. Thus, function-on-function regression models \citep{Morris.2015} are particularly appealing. %\color{red}, which -- in contrast to many time series approaches -- are not based on a predefined lag of signals, but can implicitly estimate the influence of all possible lags.\color{black}
As both the EEG and the EMG signal can be understood as noisy observations of functional variables, %i.e., of intrinsically smooth processes, 
function-on-function regression approaches offer another possibility to describe and infer the relationship of such time series \citep[see][for a recent review]{Morris.2015}. Function-on-function regression models adapt the principle of standard regression by allowing for a functional response as well as functional covariates. The so called historical model \citep{Malfait.2003, Harezlak.2007} is one possibility to explain a functional response $Y(t), t \in \mathcal{T} = [T_1, T_2]$ with $T_1, T_2 \in \mathbb{R}$, using a linear effect of the complete history of a functional covariate $X(s), s\in\mathcal{T}$: 
\begin{equation}
\mathbb{E}(Y(t)|X=x) = \int_{T_1}^t x(s)\beta(s,t)ds. \label{eq:hist1}
\end{equation}
In contrast to the existing approaches discussed above, historical models allow us to relate a given time point of one time series to more than one time point \textcolor{myc}{in $[T_1,t]$} of another time series. %Functional historical models may thus be seen as either a generalization of finite lag time series approaches without the need of specifying a certain lag or a generalization of infinite lag models \citep{Greene.2003} for data consisting of more than one (pair of) observed time series. %Additionally, the simple historical model in (\ref{hist1}) can be extended to more complex settings including multiple functional covariate effects.

%Historical models have primarily received attention in the literature in the last decade. 
Early core work on functional historical models is limited to historical models with only one functional covariate. A multitude of application possibilities are conceivable and historical models have been used in different research areas including health and biological science \citep{Malfait.2003, Harezlak.2007, Gervini.2015, Brockhaus.2016}.
%, effects of particulate matter exposure on heart rate variability \citep{Harezlak.2007}, analyses of neoromotor data \citep{Gervini.2015} or for monitoring of cell dry mass during a biotechnological fermentation \citep{Brockhaus.2016}. 
\citet{Brockhaus.2016} extended the framework of a simple historical model as in (\ref{eq:hist1}) to functional regression models with a high number of functional historical effects and potentially further covariate effects by utilizing gradient boosting for estimation. 

Alternative estimation procedures for flexible function-on-function regression models including historical effects are based on a mixed model representation \cite[cf.][]{Scheipl.2015} or component-wise gradient boosting \cite[cf.][]{Brockhaus.2015, Brockhaus.2016}. These are implemented in the \texttt{pffr}-function of the \texttt{R} package \texttt{refund} \citep{refund.2015} and in the R package \texttt{FDboost} \citep{fdboost.2016}, respectively. %Though comprising a wide class of functional regression models, it is neither possible to specify factor-specific functional (historical) effects in the \texttt{pffr}-function, nor, as far as we know, in any other software package. 
% The gradient boosting approach being particular suitable for large data sets, we use and extend the approach of \citet{Brockhaus.2015, Brockhaus.2016}.

\textit{Proposed approach}. 
% In order to detect synchronisation of both signals (EEG and EMG signal), we use function-on-function regression models for the given time series and estimate the underlying relationship by a component-wise gradient boosting algorithm. Previous attempts to investigate the synchronisation of EEG and EMG signals are based on coherence or time series analyses \citep[see e.g.][]{Cohen.2014}, which in many aspects do not provide the flexibility of function-on-function regression approaches. 
%As gradient boosting is very flexible and scales well to large datasets, using it for estimating a function-on-function regression model \citep{Brockhaus.2015} is particularly appealing for our application.%, given the large number of functional observations. 
In order to reflect the study design in this application, we extend functional historical models to historical effects that vary over one or two (penalised) categorical covariates %\emph{random historical effects}, \emph{factor-specific historical effects} as well as \emph{random and factor-specific historical effects} 
to allow for subject-, setting- and subject-by-setting-specific effects. \textcolor{myc}{We provide mathematical concepts for the construction of design matrices, penalty matrices as well as suitable identifiability constraints. We integrate these concepts into the framework of \citet{Brockhaus.2015, Brockhaus.2016} and implement them for estimation via component-wise gradient boosting. We also speed up the estimation by making use of our particular model structure.} %Fitting such a complex structure is not a common application in function-on-function regression and, to our knowledge, has no yet been considered in the literature. 
By carrying out estimation with component-wise gradient boosting \textcolor{myc}{as in \citet{Brockhaus.2015}}, our approach has several advantages. In particular, it can fit multiple factor-, subject- as well as subject-by-factor-specific functional effects, which is not possible in \textcolor{myc}{alternative} approaches for function-on-function regression such as implemented in the \texttt{pffr}-function in the R package \texttt{refund} \citep{refund.2015}. % or the \texttt{linmod}-function in the R package \texttt{fda} \citep{fda.2014}. 
Furthermore, \textcolor{myc}{the algorithm} allows for different loss functions and thus covers models beyond mean regression \citep{Kneib.2013}, e.g., median, robust or quantile regression. It can deal with high-dimensional data sets that often go hand in hand with multi-sensor bioelectrical signal data collections, as well as settings with more covariates than observations. Our approach is able to find multi-modal effect surfaces and band effects, thereby covering special cases of time series approaches. %The promising results may even promote it as a more generally applicable method for other dynamically changing (bio-)signal relationships. 
In addition, we \textcolor{myc}{%discuss identifiability constraints that yield useful interpretations of model components, 
derive options to reduce computation time as well as memory storage considerably and address the question of uncertainty in complex boosted models.}

The remainder of this paper describes the proposed model and method in section 2, presents the gradient boosting algorithm in section 3 and covers a simulation study in section 4. We apply boosted historical models to the emotion components data in section 5 and conclude with a discussion in section 6. Our proposed methods are implemented in the R package \texttt{FDboost}, an extension of the model based boosting package \texttt{mboost} \citep{mboost.2016}. The R code for our simulation as well as \textcolor{myc}{code and data} for our application is provided in an online repository (\url{https://github.com/davidruegamer/BoostingSignalSynchro}).

\section{Functional response models and historical effects}

After outlining the functional historical model in \ref{sec:histFirst}, we extend the model of \citet{Brockhaus.2016} to models with functional historical terms interacting with categorical covariates and to random functional historical effects in subsection \ref{sec:facSpecHist}.

\subsection{Functional historical models} \label{sec:histFirst}

We focus on additive functional regression models of the form \citep{Brockhaus.2015, Brockhaus.2016} 
\begin{equation}
\xi(Y(t)|\bm{X}=\bm{x}) = h(\bm{x})(t) = \sum_{j=1}^J h_j(\bm{x})(t), \label{eq:addMod}
\end{equation}
where $\xi$ is a transformation function for the conditional distribution of the functional response $Y(t), t \in \mathcal{T}$. In our application $\xi$ is equal to the conditional expectation $\mathbb{E}$, although it could also be e.g. the (pointwise) median or a quantile. The covariate set $\bm{x}$ comprises functional observations $x_1(\cdot),\ldots,x_{p_x}(\cdot)$ and scalar covariates $z_1, \ldots, z_{p_z}$ \textcolor{myc}{with $p := p_x + p_z$}. $h_j(\bm{x})(t)$ are partial effects, which can depend on scalar as well as on functional covariates. In particular, this general model class includes models with one or more historical effects 
\begin{equation}
h_j(\bm{x})(t) = \int_{l(t)}^{u(t)} x_{k_j}(s)\beta_j(s,t)ds, \label{eq:histExpl1}
\end{equation}
$k_j \in \lbrace 1,\ldots, p_x \rbrace$, which can have general integration limits $l(t)$ and $u(t)$, e.g., defined by $l(t) = T_1$ and $u(t) = t$, $l(t) = t-\delta$ and $u(t) = t$ \textcolor{myc}{or partial histories $l(t) = t-\delta_l$ and $u(t) = t - \delta_u, t > \delta_l > \delta_u  > 0$ as in \citet{Harezlak.2007}}. Functional historical effects are particularly suited to settings where both response $Y(t)$ and covariates $X_{k_j}(s)$ are observed over the same time interval, $s,t \in \mathcal{T}$. % = [T_1, T_2]$, $T_1, T_2 \in \mathbb{R}$ and $T_1 \leq T_2$. 

In practice, $x_{k_j}(\cdot)$ is observed on a grid $s_1, \ldots, s_R~$ and the integral as well as the smooth coefficient surface $\beta_j(s,t)$ in (\ref{eq:histExpl1}) must be approximated. \textcolor{myc}{We use numerical integration and a tensor product spline basis expansion, respectively.}  %, this is done as follows \citep{Scheipl.2015, Brockhaus.2016}. On the one hand, $\beta(s,t)$ can be represented using spline basis functions \citep[see e.g.][]{Wood.2006}. 
\textcolor{myc}{For $k=1,\ldots,K_x$, $l=1,\ldots,K_t$ define the basis functions $\Phi^s_{j,k}(s)$ and $\Phi^t_{j,l}(t)$ for the $s$- and the $t$-direction of the coefficient surface $\beta_j(s,t)$, respectively. Let $\theta_{j,k,l}$ be the corresponding basis coefficients and $\Delta(s_r)$ numerical integration weights for the observed time points $s_r$.} Then, 
% \begin{equation}
% \beta(s,t) \approx \sum_{k=1}^{K_x} \sum_{l=1}^{K_t} \Phi^s_{k}(s) \Phi^t_{l}(t)\, \theta_{k,l}  \label{approx1}
% \end{equation}
% with corresponding spline coefficients $\theta_{k,l}, k=1,\ldots,K_x, l=1,\ldots,K_t$ \citep{Wood.2006}. 
% On the other hand, the integral is approximated via an appropriate numerical integration scheme. Let $\Delta (s_r)$ be the corresponding numerical integration weights for the observed time points $s_r$, $r = 1,\ldots,R$. Then %, the historical effect can be approximated \citep{Scheipl.2015} via 
% \begin{equation}
% \int_{l(t)}^{u(t)} x(s) \, \beta(s,t)\, ds  \approx \sum_{r=1}^R \underbrace{[ I(l(t) \leq s_r \leq u(t)) \, \Delta(s_r) \, x(s_r) ]}_{:= \tilde x(s_r,t)} \beta(s_r,t), \label{approx2}
% \end{equation}
% \noindent where $I(\cdot)$ is the indicator function. Putting (\ref{approx1}) and (\ref{approx2}) together the historical effect is represented by
% \begin{equation}
% \int_{l(t)}^{u(t)} x(s) \beta(s,t)\, ds \approx \sum_{r=1}^R \tilde x (s_r,t) \sum_{k=1}^{K_x}\sum_{l=1}^{K_t} \Phi^s_{k}(s_r) \Phi^t_{l}(t) \, \theta_{k,l}. \label{numOfHist}
% \end{equation}
the historical effect can be represented by \citep{Scheipl.2015, Brockhaus.2016}
% \begin{equation}
% \int_{l(t)}^{u(t)} x(s) \beta(s,t)\, ds \approx \sum_{r=1}^R \underbrace{\left[\, I \! \left(\, l(t) \leq s_r \leq u(t) \, \right) \, \Delta(s_r) \, x(s_r) \, \right]}_{:= \tilde x(s_r,t)} \sum_{k=1}^{K_x}\sum_{l=1}^{K_t} \Phi^s_{k}(s_r) \Phi^t_{l}(t) \, \theta_{k,l} \label{numOfHist}
% \end{equation}
%with indicator function $I(\cdot)$. 
%Using the Kronecker-product $\otimes$, %(\ref{numOfHist}) can alternatively represented as 
\begin{equation}
\int_{l(t)}^{u(t)} x_{k_j}(s) \beta_j(s,t)\, ds \approx \bm{B}_j(x_{k_j},t) \bm{\theta}_j\label{eq:numOfHist2}
\end{equation}

\noindent with $\bm{\theta}_j = (\theta_{j,1,1}, \ldots, \theta_{j,K_x,K_t})^\top$, $\bm{B}_j(x_{k_j},t) = \bm{B}_j^x(x_{k_j},t) \otimes \bm{B}_j^t(t)$ using the Kronecker-product $\otimes$ and by defining
\begin{equation*}
\bm{B}_j^x(x_{k_j},t) =  \left[\sum_{r=1}^R \Delta(s_r) \, x_{k_j}(s_r, t) \Phi^s_{j,1}(s_r)\ \cdots\ \sum_{r=1}^R \Delta(s_r) \, x_{k_j}(s_r,t) \Phi^s_{j,K_x}(s_r)\right]
\end{equation*}
as well as $\bm{B}_j^t(t) = [\Phi^t_{j,1}(t)\ \cdots\ \Phi^t_{j,K_t}(t)]$. Let $I(\cdot)$ be the indicator function. Following \citet{Scheipl.2015}, for $n$ observed curves $x_{k_j,1}(\cdot), \ldots, x_{k_j,n}(\cdot)$ at grid points $s_r$, $x_{k_j}(s_r, t) = x_{k_j}(s_r) \cdot I \lbrace l(t) \leq s_r \leq u(t) \rbrace$ and response observations $y_i(t_{i,d})$ at potentially curve specific time points $t_{i,d} \in \mathcal{T}, i=1,\ldots,n, d=1,\ldots,D_i, N = \sum_{i=1}^n D_i$, the design matrix of a historical effect can be summarised by 
\begin{equation}
\mathcal{B}_j := \bm{B}_j^x \odot \bm{B}_j^t = (\bm{B}_j^x \otimes \bm{1}^\top_{K_t}) \ast (\bm{1}^\top_{K_x} \otimes \bm{B}_j^t), \label{eq:desmatHist}
\end{equation}
where $\bm{B}_j^x \in \mathbb{R}^{N \times K_x}$ with rows $\bm{B}_j^x(x_{k_j,i},t_{i,d})$, $\bm{B}_j^t \in \mathbb{R}^{N \times K_t}$ with rows $\bm{B}_j^t(t_{i,d})$, $\odot$ is the row-wise tensor product, $\ast$ the Hadamard product (element-wise matrix multiplication) and $\bm{1}^\top_a$ a row-vector of length $a$. In the supplemental material, we provide a simple example of how to interpret estimated coefficient surfaces of historical effects, as we believe that this is an \textcolor{myc}{important} part in using historical models.

Regularisation of the coefficient vector $\bm{\theta}_j$ in (\ref{eq:numOfHist2}) is achieved by an anisotropic penalty. Using the marginal penalties $\bm{P}_j^x \in \mathbb{R}^{K_x \times K_x}$ and $\bm{P}_j^t \in \mathbb{R}^{K_t \times K_t}$ of the historical effect basis in $s$- and $t$-direction, respectively, a quadratic penalty term can be constructed as
\begin{equation}
\bm{\theta}_j^\top \, \mathcal{P}_j \, \bm{\theta}_j = \bm{\theta}_j^\top \left[ \lambda_j^{x} (\bm{P}_j^x \otimes \bm{I}_{K_t}) + \lambda_j^{t} (\bm{I}_{K_x} \otimes \bm{P}_j^t) \right] \bm{\theta}_j = \bm{\theta}_j^\top \left[ \lambda_j^{x} \bm{P}_j^x  \oplus \lambda_j^t \bm{P}_j^t \right] \bm{\theta}_j, \label{eq:pen}
\end{equation}
where $\lambda_j^{x}, \lambda_j^{t}\geq 0$ are smoothing parameters and $\oplus$ is the Kronecker-sum \citep{Wood.2006, Scheipl.2015}. More details on the penalisation and potential extensions can be found in the next subsection. Similarly, penalised basis expansions as in (\ref{eq:numOfHist2}), (\ref{eq:desmatHist}), and (\ref{eq:pen}) can also be constructed for a multitude of other effects of scalar and / or functional covariates, including all effects of scalar covariates in our proposed model for the emotion components data \citep{Scheipl.2015, Brockhaus.2015}. 

In addition to ordinary historical effects, this approach can incorporate a time-varying intercept $h_j(\bm{x})(t)=\alpha(t)$ as well as time-varying categorical or random effects 
\begin{equation}
h_{j}(\bm{x})(t) = \gamma_{j,e}(t) \cdot I(z_{q_j} = e), \label{eq:catEff}
\end{equation}
where $q_j \in \lbrace 1, \ldots, p_z \rbrace$, $z_{q_j}$ is a categorical covariate with levels $e \in \lbrace 1,\ldots,\eta \rbrace$ and $\gamma_{j,e}(t)$ the corresponding time-varying coefficient. % and $C$ a contrast of $z_{q_j}$. In the following, we assume categorical covariates to be either dummy coded, $C(z_{q_j},e) \equiv I(z_{q_j}=e)$, or effect coded, $C(z_{q_j},e) \equiv \psi(z_{q_j},e)$. In particular, $C$ is equivalent to a dummy coding for time varying random effects. 
The smoothness of the coefficient functions $\alpha(t)$ and $\gamma_{j,e}(t)$ is obtained with a spline basis representation  
%with $\bm{B}^x_j(x,t) = 1 \otimes \bm{B}_j^t(t)$ respectively $\bm{B}(\bm{z}_{q_j}) \otimes \bm{B}_j^t(t)$ instead of 
as in (\ref{eq:numOfHist2}) and a Kronecker sum penalty as in (\ref{eq:pen}) with %with $\bm{P}_j^t$ a smoothness penalty matrix, 
$\bm{P}_j^x$ set to zero for categorical effects and $\bm{P}_j^x = \bm{I}_{K_z}$ for (independent) functional random effects \citep[see][for more details]{Brockhaus.2015}. %If $Y(t)$ follows a normal distribution with variance $\sigma_\varepsilon^2$, 
In particular, for functional random effects, the quadratic penalty in (\ref{eq:pen}) is equivalent to a normal distribution assumption on $\bm{\theta}_j$ with zero mean and covariance proportional to the generalised inverse of $\mathcal{P}_j$ \citep{Brumback.1999}, inducing a Gaussian process assumption for the functional random effects. Furthermore, we consider interaction effects of $z_{q_j}$ and a second categorical covariate $z_{q^\prime_{j}}$ with levels $f = 1,\ldots, \varphi$ of the form 
\begin{equation}
h_j(\bm{x})(t) = \rho_{j,e,f}(t) \cdot I(z_{q_j} = e) \cdot I(z_{q^\prime_{j}} = f). \label{eq:iaEff}
\end{equation}
Identifiability constraints for time-varying categorical effects such as (\ref{eq:catEff}) and (\ref{eq:iaEff}) are discussed in the following subsection.
% Apart from the previous presented historical effects, we pay particular attention to factor-specific historical effects, which represent a completely new approach in functional regression modeling. Accompanied with even more entangled challenges especially regarding the correct specification and implementation when including such effects in our models, we separately cover some theory about factor-specific historical effects in the next subsection as well as emphasising the corresponding key issues in the implementation in \ref{grabo}. 

\subsection{Factor-specific historical effects} \label{sec:facSpecHist}

In light of our application, we \textcolor{myc}{newly} introduce factor-specific historical effects for functional regression models. Factor-specific historical effects can be useful when historical effects are assumed to vary, e.g., between different study settings or subjects. First, consider a categorical covariate ${z}_{q_j}$ with levels $e=1,2,\ldots,\eta$ and a functional covariate ${x}_{k_{j}}(s)$, which is modeled via a historical effect. A simple additive model of the form (\ref{eq:addMod}) would then include a main historical effect (\ref{eq:histExpl1}) and a factor-specific historical effect 
\begin{equation}
% \begin{split}
h_j(\bm{x})(t) = I ({z}_{q_j} = e) \cdot \int_{l(t)}^{u(t)} {x}_{k_j}(s)\beta_{j,e}(s,t)ds. \label{simpleHistEff}
% \end{split}
\end{equation}
% for an effect coded categorical covariate or  
% \begin{eqnarray}
% \mathbb{E}({Y}(t)|\bm{X}=\bm{x}) & = & I({z}_{q_j} = e) \cdot \frac{1}{u(t)-l(t)} \int_{l(t)}^{u(t)} {x}_{k_j}(s)\beta_{k_j,e}(s,t)ds \label{simpleHistDummy}
% \end{eqnarray}
% using a dummy coding for $z_{q_j}$.
% Figure ... shows an example of the estimated coefficient surfaces $\beta_j(s,t)$ and $\beta_{k_j,e}(s,t)$, $e=1,2$ along with the corresponding functional observations.\\
% 
% FIGURE\\
% % \begin{figure}
% % \centering
% % \makebox{\includegraphics{figure.eps}}
% % \caption{\label{fig01}A figure for test.}
% % \end{figure}
Given a total of $N$ observations and the covariate vector $\bm{z}_{q_j} = \left[ ({z}_{q_j,1} \otimes \bm{1}_{D_1})^\top, \ldots, ({z}_{q_j,n} \otimes \bm{1}_{D_n})^\top \right]^\top$ the factor-specific historical effect is constructed similarly to (\ref{eq:desmatHist}). The design matrix is extended to 
\begin{equation}
\mathcal{B}_j = \bm{B}_j^z(\bm{z}_{q_j}) \odot \bm{B}_j^x \odot \bm{B}_j^t = \widetilde{\bm{B}}_j^x \odot \bm{B}_j^t, \label{eq:desMat1}
\end{equation}
where $\bm{B}_j^z(\bm{z}_{q_j})$ is a design matrix for the factor variable depending on the constraints on $\beta_{j,e}(s,t)$ (see below) and $\widetilde{\bm{B}}_j^x = \bm{B}_j^z(\bm{z}_{q_j}) \odot \bm{B}_j^x$. An important special case is given for the unconstrained estimation of $\beta_{j,e}$ when the observations are sorted by the factor levels $e = 1,\ldots,\eta$. This yields a block-diagonal incidence matrix for $\bm{B}_j^z(\bm{z}_{q_j}) = \text{diag} (\bm{1}_{\kappa_1}, \bm{1}_{\kappa_2}, \ldots, \bm{1}_{\kappa_\eta} ) \in \mathbb{R}^{N \times \eta}$ and a $N \times \eta K_x$ block-diagonal matrix for 
$\widetilde{\bm{B}}_j^x = \text{diag} (\bm{B}^x_{j,1}, \ldots, \bm{B}^x_{j,\eta} )$.  
% \begin{equation*}
% \left( \,\begin{array}{r r c r}
% \cline{1-1}
%  \mclr{1}   \\
%  \mclr{\vdots}  \\
%  \mclr{1} \\
% \cline{1-2}
%  & \mclr{1}  \\
%  & \mclr{\vdots}  \\
%  & \mclr{1} \\
% \cline{2-2} 
%  & & \ddots\\
% \cline{4-4}
% & & & \mclr{1}  \\
% & & & \mclr{\vdots}  \\
% & & & \mclr{1} \\
% \cline{4-4}
% \end{array}\,\right)
% \end{equation*}
% $$\bm{B}^z(\bm{z}_{q_j}) = 
% %\begin{pmatrix} 
% \text{diag}\left(\bm{1}_{\kappa_1}, 
% %& & & \\ & 
% \bm{1}_{\kappa_2}, %& & \\ & & \ddots & \\ & & & 
% \ldots, \bm{1}_{\kappa_\eta} \right),% \end{pmatrix} 
% %\quad , 
% \qquad \widetilde{\bm{B}}^x = 
% %\begin{pmatrix} 
% \text{diag}\left(\bm{B}^x_{\kappa_1},
% %& & & \\ & 
% \bm{B}^x_{\kappa_2}, %& & \\ & & \ddots & \\ & & & 
% \ldots, \bm{B}^x_{\kappa_\eta} \right) % \end{pmatrix}
% $$ 
Here, $\bm{B}^x_{j,e} \in \mathbb{R}^{\kappa_e \times K_x}$ contains the rows $\sum_{k=1}^{e-1} \kappa_k + 1,\ldots,\sum_{k=1}^{e} \kappa_k$ of $\bm{B}_j^x$ corresponding to all rows with factor level $e$ and $\kappa_e$ being the total number of observation points for factor level $e$.  % $\bm{B}^x$ and $\bm{B}^t$ therefore must be constructed by appending $\bm{B}^x(x_i,t_{i,d})^\top$ respectively $\bm{b}^t(t_{i,d})^\top$ in the order of sorted observations, i.e. $\bm{B}^t = \left(\bm{b}^t(t_{(1),1}), \bm{b}^t(t_{(1),2}), \ldots, \bm{b}^t(t_{(1),D_{(1)}}), \bm{b}^t(t_{(2),1}), \ldots, \bm{b}^t(t_{(N),D_{(N)}}) \right)^\top$ and analogously for $\bm{B}^x$.
This special structure can be exploited for a more efficient computational implementation (see section \ref{impSteps} for more details).

%An additional challenge arises 
When the historical effect of ${x}_{k_j}$ is not only factor- or subject-specific, but varies for a categorical covariate ${z}_{q_j}$ with levels $e=1,2,\ldots,\eta$ as well as for subjects ${z}_{q^\prime_{j}}$ with levels $f = 1,2,\ldots,\varphi$, we let
\begin{equation}
h_j(\bm{x})(t) = I({z}_{q_j} = e) \cdot I({z}_{q^\prime_{j}} = f) \cdot  \int_{l(t)}^{u(t)} {x}_{k_j}(s)\beta_{j,e,f}(s,t)ds. \label{eq:doubvarhist}
\end{equation}
The design matrix for the random factor-specific historical effect or \emph{doubly-varying historical effect} (\ref{eq:doubvarhist}) is then defined by extending $\bm{B}_j^z(\bm{z}_{q_j})$ in (\ref{eq:desMat1}) to
\begin{equation*}
\bm{B}_j^z(\bm{z}_{q_j}, \bm{z}_{q^\prime_{j}}) = \bm{B}_j^z(\bm{z}_{q_j}) \odot \bm{B}_j^z(\bm{z}_{q^\prime_{j}}). %\label{eq:doubleDesMat}
\end{equation*}
For these factor-specific historical effects (\ref{simpleHistEff}) and (\ref{eq:doubvarhist}), we have to carefully consider their identifiability and regularisation.

\textit{Identifiability constraints}. In order to ensure that the main historical effect %$$\frac{1}{u(t)-l(t)} \int_{l(t)}^{u(t)} {x}_{k_j}(s)\beta_j(s,t)ds$$ 
is separable from the factor-specific historical effects and vice versa, we impose the following constraint when both are included in the model:
\begin{equation}
\sum_{e=1}^\eta \psi_e \cdot \beta_{j,e}(s,t) = 0 \quad \forall t \in \mathcal{T}, s\in [l(t),u(t)], \label{sumToZeroConstr}
\end{equation}
where $\psi_e$ are weights for each level $e=1,\ldots,\eta$ of the factor variable. Specifically, for observed curves $i=1,\ldots,n$, we use $\psi_e  = \sum_{i=1}^n I(z_{q_j,i}=e)$, which coincides with equal weighting in the case of balanced factor levels.
This also allows $\beta_j(s,t)$ in (\ref{simpleHistEff}) to be interpretable as average historical effect over the $\eta$ subgroups. (\ref{sumToZeroConstr}) ensures identifiability because the factor-specific historical effects are centred around the surface of the main effect for models including \textcolor{myc}{both} (\ref{eq:histExpl1}) and (\ref{simpleHistEff}). 
% In turn, for the main historical effect to be separable from the time-varying intercept, we impose the constraint
% \begin{equation}
% \int_{l(t)}^{u(t)} x_{k_j,i} \beta_{k_j}(s,t)\,ds = 0 \quad \forall t\in \mathcal{T} \quad \forall i \in \lbrace 1,\ldots,N \rbrace \label{sumToZeroConstr0}
% \end{equation}
% for every observation $i$.

For the doubly-varying historical effects to be defined as deviations from both factor-specific historical effects, we impose the constraints 
\begin{eqnarray}
\sum_{e = 1}^\eta \psi_{e,f} \cdot \beta_{j,e,f}(s,t) \!\!\!\!&=&\!\!\! 0 \,\,\,\forall\,t \in \mathcal{T}, s\in [l(t),u(t)], \, f \in \lbrace 1,\ldots,\varphi \rbrace \label{sumToZeroConstr2} \,\,\, \text{and} \\
\sum_{f = 1}^\varphi \psi_{e,f} \cdot \beta_{j,e,f}(s,t) \!\!\!\!&=&\!\!\! 0 \,\,\,\forall\,t \in \mathcal{T}, s\in [l(t),u(t)], \, e \in \lbrace 1,\ldots,\eta \rbrace, \label{sumToZeroConstr3}
\end{eqnarray}
for which we use the weights $\psi_{e,f} = \sum_{i=1}^n I(z_{q_j,i} = e, z_{q^\prime_{j},i} = f)$. 

To ensure identifiability and interpretability of the whole model, further constraints must be placed on effects other than the historical effects, i.e., when including time-varying effects in the model. As in \citet{Scheipl.2015} and \citet{Brockhaus.2015}, all time-varying effects in our models are specified as deviations from the smooth intercept $\alpha(t)$. This ensures the identifiability of each effect and allows for a meaningful interpretation (as deviation from the sample mean $\alpha(t)$). Consider the factor variable $z_{q_j}$ and an effect as in (\ref{eq:catEff}). We then impose 
%\begin{equation*}
$\sum_{e=1}^{\eta} \psi_e \cdot \gamma_{j,e}(t) = 0 \, \forall \,t\in \mathcal{T}$. %\label{facOne}
%\end{equation*}
A similar constraint is enforced for interaction effects (\ref{eq:iaEff}) with coefficients $\rho_{j,e,f}(t)$: %\begin{eqnarray*}
$\sum_{e=1}^{\eta} \psi_{e,f} \cdot \rho_{j,e,f} (t) = 0 \,\forall \,t\in \mathcal{T}, \,f\in\lbrace 1,\ldots,\varphi\rbrace$ and % \quad\text{and}\\
$\sum_{f=1}^{\varphi} \psi_{e,f} \cdot \rho_{j,e,f} (t) = 0 \, \forall \,t\in \mathcal{T}, \,e\in\lbrace 1,\ldots,\eta\rbrace$,
%\end{eqnarray*}
i.e., each interaction effect has to be centred around its corresponding main effects. For details on the implementation, see section B in the supplementary material.

\textit{Parameterisation}. %As for interactions of covariates in non-functional models, we assume in the following that (a) a factor-specific historical effect with a categorical covariate in dummy coding and constraint (\ref{sumToZeroConstr}) is always accompanied by the corresponding main effects (an ordinary historical effect and a time-varying factor effect) and that (b) a doubly-varying historical effect with constraints (\ref{sumToZeroConstr2}), (\ref{sumToZeroConstr3}) is always included in a model along with the single-varying historical effects and the two-way interaction effect of the factors interacting with the historical effect. This formulation is particularly useful in the light of model selection, facilitating a data-driven choice on whether the historical effect interacts with the factor variable in the first place.
The separation of the factor-specific historical effect and the corresponding main historical effect together with constraint (\ref{sumToZeroConstr}) is particularly useful in the light of model selection. However, an alternative model formulation that does not separte main and factor-specific historical effects may sometimes be beneficial for the interpretation of estimated effects and the simplicity of the model definition. 
% A different parameterisation ... Given $n$ observations $x_{i,j}(t), i=1,\ldots,n$, for $t\in[T_1,T_2]$, any (factor-specific) historical effect of the form $$I(z=e) \cdot \int_{l(t)}^{u(t)} x_{j,i}(s) \beta_{k_j,e} (s,t)\, ds$$ can be rewritten as 
% \begin{equation}
% I(z=e) \cdot \int_{l(t)}^{u(t)} \left[ x^c_{j,i}(s) + \bar{x}_j(s) \right] \beta_{k_j,e} (s,t)\, ds, \label{meancent}
% \end{equation}
% with the centred covariate $x^c_{j,i}(s) = x_{j,i}(s) - \frac{1}{N} \sum_{i=1}^N x_{j,i}(s) =: x_{j,i}(s) - \bar{x}_j(s)$. The integral in (\ref{meancent}) can be split up into the sum of two terms. The second part $\int_{l(t)}^{u(t)} \bar{x}_j(s) \beta_{k_j,e} (s,t)\, ds$ being independent of $i$, it can be alternatively defined as $\zeta_e(t)$ by integrating over $s$, which can analogously be done for an ordinary historical effect with $\zeta(t)$ as corresponding result. 
A historical model with a main and factor-specific historical effects can be rewritten as 
% follows:
% \begin{equation}
% \begin{split}
% \xi(Y_i(t)|\bm{X}_i=\bm{x}_i) &= \int_{l(t)}^{u(t)} {x}_{k_j}(s)\beta_j(s,t)ds + I ({z}_{q_j} = e) \cdot %\frac{1}{u(t)-l(t)} 
% \int_{l(t)}^{u(t)} {x}_{k_j}(s)\beta_{j^\prime,e}(s,t)ds\\
%&= 
$\int_{l(t)}^{u(t)} {x}_{k_j}(s) \Bigl( \beta_{j}(s,t) + I({z}_{q_j} = e) \!\cdot \! \beta_{j,e}(s,t) \Bigr)\, ds$, 
%.\label{eq:interp}
%\end{split}
%\end{equation}
%by defining $\tilde{\alpha}(t) := \alpha(t) + \zeta(t)$ and $\tilde{\gamma}_{0,e}(t) := \gamma_{0,e}(t) + \zeta_e(t)$. 
combining main and factor-specific historical effects by estimating the sum $\tilde{\beta}_{j,e}(s,t) := \left(\beta_{j}(s,t) +  I({z}_{q_j}=e) \cdot \beta_{j,e}(s,t)\right)$ and thereby making constraint (\ref{sumToZeroConstr}) obsolete.

\textit{Regularisation}. For the regularisation of a factor-specific historical effect, the penalty depends on whether we want to regularise over the factor levels, e.g., for ``\textit{random historical effects}'', or not, e.g., for study settings. In general, the quadratic penalty matrix in (\ref{eq:pen}) is extended to an anisotropic penalty
\begin{equation}
\mathcal{P}_j = \bigg( \lambda_j^z \bm{P}_j^z \oplus \bigg[ \lambda_j^x \bm{P}_j^x \oplus \lambda_j^t \bm{P}_j^t \bigg] \bigg), 
\label{pen2way}
\end{equation}
where $\bm{P}_j^z$ is the $K_z \times K_z$ marginal penalty matrix over the factor levels and $\lambda_j^x, \lambda_j^t, \lambda_j^z$ are the smoothing parameters controlling the regularisation of the historical effect part in $s$- as well as $t$-direction and of the factor variable part, respectively. Usually, $K_z$ is the number of factor levels (minus one, depending on the constraint on the effect) and $\bm{P}_j^z$ is a simple Ridge penalty $\bm{P}_j^z = \bm{I}_{K_z}$. Whereas the factor-specific historical effect is therefore \textcolor{myc}{shrunk} towards the main historical effect in a model with both, main and factor-specific historical effect, the penalty in the alternative parametrisation without constraint on the factor-specific historical effect enforces shrinkage of $\beta_{j,e}$ towards zero. In practice, the $s$- and $t$-directions of the historical effect are typically measured on the same scale (i.e., time), thus we introduce an isotropic penalty for the historical effect part by defining $\lambda_j^t \equiv \lambda_j^x =: \lambda_j^h$ and $\bm{P}_j^x \oplus \bm{P}_j^t =: \mathcal{P}_j^h$. % Regarding the two different parameterisations of a model with main and factor-specific historical effects, regularisation the penalty leads to slightly different effects.
For the doubly-varying historical effect (\ref{eq:doubvarhist}), the term $\lambda_j^z \bm{P}_j^z$ in (\ref{pen2way}) is replaced by $\left[ \lambda_j^{z} \bm{P}_j^{z} \oplus \lambda_j^{z'} \bm{P}_j^{z'} \right]$. If one or both factors are not penalised, the corresponding penalty matrices are set to zero. %If both factors are penalised with a Ridge penalty, $\mathcal{P}_j  = (\lambda_j^z + \lambda_j^{z'}) \bm{I}_{K_z K_{z'}} \oplus \lambda_j^h \mathcal{P}_j^h$.

\section{Estimation: Component-wise gradient boosting} \label{grabo}

The estimation via component-wise gradient boosting \citep{Buehlmann.2007,Brockhaus.2015} has several advantages. %Due to its generic framework, it does not require additional fitting routines for more complicated model formulations, only an extension of the set of model terms. 
The main advantage of using component-wise boosting over conventional estimation procedures lies in the nature of component-wise fitting, %Whereas conventional approaches suffer from ever-growing run times the more complex models become,
as the feasibility of component-wise fitting procedures only depends on the most complex individual component. Adding partial effects step-by-step, boosting provides implicit variable selection and allows for model estimation in settings with $J > n$ or $p > n$. %This is particularly favourable for models with a large number of additive components.%\\
% \noindent We now present the algorithm and additionally highlight important steps and challenges in its application. 

\subsection{Component-wise gradient boosting}

The component-wise gradient boosting algorithm for a function-on-function regression model was introduced by \citet{Brockhaus.2015}, and is based on the functional gradient descent (FGD) algorithm \citep[cf.][]{Buehlmann.2007, mboost.2016}.

\textit{Loss function and empirical risk}. In general, the com\-po\-nent-wise FGD algorithm aims to minimize the expected loss $\mathbb{E}_{(Y,\bm{X})}(\rho(({Y},\bm{X}),h))$ for response $Y$ and covariates $\bm{X}$ with respect to the additive predictor $h$ for a suitable loss function $\rho$. The loss is determined by the underlying regression problem, e.g., the$~L^2$-loss for mean regression. %For functional observations, let $\rho:(\mathcal{Y} \times \mathcal{X}) \times \mathcal{H} \to L^1(\mathcal{T},\mu)$ be a loss function measuring the loss at each $t$, differentiable with respect to $h$, let $\mu$ be the Lebesgue measure and $\mathcal{H}$ the set of all functions from $(\mathcal{X}\times\mathcal{T})$ to $\mathcal{Y}$. 
In order to adapt the principle of FGD to functional observations, the loss function $\ell$ for a whole trajectory is defined as %$\ell: (\mathcal{Y} \times \mathcal{X}) \times \mathcal{H} \to \mathbb{R}^{+}_0; \,
$\ell((Y,\bm{X}),h) = \int_\mathcal{T} \rho((Y,\bm{X}),h)(t)\,dt$, i.e., the integrated pointwise loss $\rho$ over the domain $\mathcal{T}$.
For potentially functional observations $(y_i, \bm{x}_i), i=1,\ldots,n$, the objective function, the risk, is then given by $\mathbb{E}_{(Y,\bm{X})}(\ell(({Y},\bm{X}),h))$ and the FGD algorithm for functional regression models aims at minimizing the empirical risk
\begin{equation*}
n^{-1} \sum_{i=1}^n \sum_{d=1}^{D_i} w_i \Upsilon(t_{i,d}) \rho((y_i, \bm{x}_i),h)(t_{i,d}), \label{eq:emprisk}
\end{equation*}
where sampling weights $w_{i}$ are used to select or deselect all observations of one functional trajectory in resampling approaches and $\Upsilon(t)$ are weights of a numerical integration scheme used to approximate the integrated loss $\ell$ \citep{Brockhaus.2016} .

\textit{Routine and baselearners}. In each step, the FGD algorithm % is an iterative procedure, which 
evaluates a set of \emph{baselearners} \textcolor{myc}{(in this case corresponding to penalised regression for the partial effects $h_j$), chooses the baselearner that best fits the negative gradient at the current estimate $-\frac{\partial}{\partial\,h} \mathbb{E}_{(Y,\bm{X})}(\ell(({Y},\bm{X}),h))$ and updates the fit in light of this choice.} As in representation (\ref{eq:numOfHist2}), we assume that every baselearner can be represented as a linear effect in $\bm{\theta}_j \in \mathbb{R}^{K_j}$, i.e. $h_j(\bm{x})(t) = \bm{B}_j(\bm{x}_{k_j},t)\bm{\theta}_j$, with suitable penalty, e.g., (\ref{eq:pen}) or (\ref{pen2way}). %A commonly used application of the FGD algorithm are structured additive regression models in the form of $$\mathbb{E}(Y_i|\bm{X_i}=\bm{x}_i) = h(\bm{x}_i) = \sum_{j=1}^J h_j(\bm{x}_i)$$ with $\bm{x}_i = (x_{1,i}, \ldots, x_{p,i})$ and predefined functions $h_j$ representing the baselearners. $h_j(\cdot)$ are possibly different modeling alternatives in the framework of structured additive regression such as a linear effect $h_j(\bm{x}_i) = x_{k_j,i}\,\beta$, $k_j\in\lbrace 1,\ldots, p \rbrace$, a categorical effect or for example a smooth effect $h_j (\bm{x}_i) = f(x_{k_j,i})$. In the stepwise procedure of FGD, one of these baselearners is iteratively added to the linear predictor. For additive functional regression models as covered in this paper, time-varying effects for categorical covariates, historical effects and factor-specific historical effects are used as baselearners. These standalone base procedures can be build up prior to the iterative process and are then evaluated in each boosting step with respect to the negative gradient of a predefined loss function. The build up for functional effects is done by using an index for the time $t$, a corresponding design matrix $B_j(t)$ as well as a penalty matrix $\mathcal{P}_j$. The smoothing parameter is obtained prior to the model fit on the basis of chosen degrees of freedom for each regularised baselearner. In \ref{impSteps} this is described in greater detail. 

\textit{Algorithm}. %Combining these weights with the weights of the integration scheme $\delta the empirical risk is 
% The empirical risk is 
% \begin{equation}
% r^\ast = \underset{h}{\text{argmin}}\quad n^{-1} \sum_{i=1}^n \sum_{d=1}^{D_i} w_i \, \tilde \rho((y_i, x_i),h)(t_{i,d}) .
% \end{equation}
The full algorithm is given by the following five steps:
% 
% \emph{Algorithm: Component-wise gradient boosting}

\begin{itemize}
\item[\emph{Step 1}:] Set $m=0$; Initialise the estimates, e.g. $\hat{\bm{\theta}}_j^{[m]} \equiv \bm{0}$ for each baselearner $j\in \lbrace 1,\ldots,J \rbrace$, and define $\hat{h}^{[m]}(\bm{x})(t) = \sum_{j=1}^J \bm{B}_j(\bm{x},t) \hat{\bm{\theta}}_j^{[m]}\,$; choose a step-length $\nu \in (0,1]$ and a maximal stopping iteration $m_{\text{stop}}$.
\item[\emph{Step 2}:] Compute the negative gradient $-\frac{\partial}{\partial\,h} \rho({(y,\bm{x})},h)$ and define the \textcolor{myc}{so called} \emph{pseudo residuals} $$u_i(t_{i,d}) := \left. -\frac{\partial}{\partial\,h}  \rho((y_i,\bm{x}_i),h)(t_{i,d}) \right|_{h=\hat{h}^{[m]}}.$$ 
%$u_i(t_{i,d})$ 
%for $i=1,\ldots,n$.% by %$$u_i(t_{i,d}) := \left. -\frac{\partial}{\partial\,h}  \rho((Y_i,X_i),h)(t_{i,d}) \right|_{h=\hat{h}^{[m]}(x_i)(t_{i,d})}.$$
\item[\emph{Step 3}:] Fit the baselearners $j=1,\ldots,J$ to the pseudo residuals $$\hat{\bm{\vartheta}}_j =  \underset{\bm{\vartheta}\in\mathbb{R}^{K_j}}{\text{argmin}} \sum_{i=1}^n \sum_{d=1}^{D_i}  w_i \Upsilon(t_{i,d}) \left\{ u_i(t_{i,d}) - \bm{B}_j(\bm{x}_{k_j,i},t_{i,d}) \bm{\vartheta} \right\}^2 + \bm{\vartheta}^\top \mathcal{P}_j \bm{\vartheta}$$ and find the best-fitting $j^\ast$th baselearner such that $$j^\ast = \underset{j=1,\ldots,J}{\text{argmin}} \sum_{i=1}^n \sum_{d=1}^{D_i}  w_i \Upsilon(t_{i,d}) \left\{ u_i(t_{i,d}) - \bm{B}_j(\bm{x}_{k_j,i},t_{i,d}) \hat{\bm{\vartheta}}_j \right\}^2.$$
\item[\emph{Step 4}:] Set $\hat{\bm{\theta}}_{j^\ast}^{[m+1]} = \hat{\bm{\theta}}_{j^\ast}^{[m]} + \nu \hat{\bm{\vartheta}}_{j^\ast}$, $\hat{\bm{\theta}}_{j}^{[m+1]} = \hat{\bm{\theta}}_{j}^{[m]} \, \forall\, j\neq j^\ast$ and update $\hat{h}^{[m]}$ accordingly.
\item[\emph{Step 5}:] Set $m = m + 1$; as long as $m \leq m_{\text{stop}}$, repeat steps 2 --- 5.
\end{itemize}
%
%This approach is implemented in the R package \texttt{FDboost}. Factor-specific functional historical effects varying over up to two categorical covariates have been newly added and can now be estimated. %This also includes the case, in which the historical model is subject- and category-specific, i.e. where one of the factors is included with a random historical effect. %The covariate used in the historical effect should be centred for performance reasons. This only changes the intercept $\alpha(t)$ to lie at the center ot the data.
%\\ %Based on the generic implementation of the R package \texttt{mboost} \citep{mboost.2016}, the R package \texttt{FDboost} allows for the specification of function-on-function regression models in the same manner as \texttt{mboost}. % for models without functional effects. 
%Additionally, \texttt{FDboost} provides a specific baselearner function for historical effects varying with regard to other categorical covariates including random effects.
%\noindent\emph{Tuning parameter}. 
The final model \textcolor{myc}{with corresponding parameters $\hat{\bm{\theta}}_j^{m^\ast}$, $j=1,\ldots,J$, $m^\ast \in \lbrace 1,\ldots,m_{\text{stop}} \rbrace$} is chosen from the set of $m_{\text{stop}}$ estimated models via cross-validation or other resampling methods on the level of curves \citep{Brockhaus.2015} \textcolor{myc}{in order to prevent over-fitting}. \textcolor{myc}{This so called} early stopping of the boosting procedure introduces regularisation on coefficient estimates \citep{Zhang.2005}.% or model selection criteria such as the BIC \citep[e.g.,][]{Buehlmann.2006}. We additionally investigate the behaviour of resampling methods and selection criteria for functional historical models, since model selection criteria in component-wise gradient boosting had so far only been considered in simple additive regression models. Since the selection criteria are based on the hat matrix of the boosting algorithm, which is calculated as $m_{\text{stop}}$ products of $N \times N$ matrices, computation time of such criteria is particularly expensive for functional data analysis and even in small settings slower than resampling methods. Yielding no superior performance in any setting, we therefore recommend to use resampling methods for model selection.

\subsection{Unbiased baselearner selection and smoothing parameter computation} \label{impSteps}

It is important to set equal degrees of freedom $df_j$ for every baselearner $j$ for a fair selection of baselearners \citep{Hofner.2011}. A regularisation over factor levels for categorical covariates with a moderate or large number of factor levels is thus often necessary in practice as $df_j$ would otherwise become very large. The smoothing parameters $\lambda_j$, which have a one-to-one correspondence with $df_j$, must therefore be computed and fixed appropriately beforehand for $j=1,\ldots,J$. Model complexity and smoothness is then controlled for fixed $\nu$ by the stopping iteration, which is chosen by resampling. 

The \texttt{FDboost} package, based on the \texttt{mboost} package, uses the Demmler-Reinsch orthogonalization \citep[DRO, see, e.g.,][]{Ruppert.2003}, which avoids repeated matrix inversions to efficiently find a suitable $\lambda_j$. %in the hat matrix %calculation for the $j$th baselearner 
%for different $\lambda_j$-values %In contrast, the \texttt{fda} package iteratively computes the inverse, which is rather inefficient. 
Nonetheless, computing the DRO may be very expensive, particularly for factor- and subject-specific historical effects, due to a singular value decomposition (SVD), and can take up to $99$\% of total computing time. To tackle this problem, on the one hand, we recommend reducing the number of knots for (doubly-)varying historical effects to a small number (e.g., four), if this is not expected to lead to unwanted oversmoothing. On the other hand, we exploit the model structure for factor-specific historical effects and derive a presentation that allows for a blockwise SVD with computation time on the order of an ordinary historical effect. This reduces overall computation time dramatically (see section C in the supplementary material for more details). For the application in section \ref{application}, for example, the most complex model with partially aggregated data could be fitted in under 16 minutes with less than 45 gigabyte RAM, whereas the brute-force method \textcolor{myc}{(fitting the model with ten knots without exploitation of the model structure)} failed, exceeding the memory limit of 1 terabyte RAM after running for more than 10 days. \textcolor{myc}{Although the first approach can be a good (approximate) ad-hoc solution, the second approach is exact and thus generally recommended if feasible.}
% Unfortunately, more elegant ways to solve this problem by for example partition the matrix $\mathfrak{Y}$ and its SVD do not seem to be feasible due to the special structure of $\mathfrak{Y}$, particularly caused by the Hadamard product in the construction of the design matrix.
% 
% \noindent \textit{Anisotropic penalties}. In contrast to the regularisation options in \texttt{mboost}, \texttt{FDboost} allows for anisotropic penalties with different smoothing parameters $\lambda^x$ and $\lambda^z$ for time-varying categorical effects. This feature plays an important role and may improve estimation performance. Depending on the particular application, it could make sense to either allow more degrees of freedom in the time direction or in the variation between the categories.

\subsection{Quantification of uncertainty}

%Several established methods for the assessment of model uncertainty and model diagnosis have been proposed for boosted models. 
\textcolor{myc}{Due to the large fluctuation in bioelectrical signals, a very important aspect in the analysis of such signals is the assessment and quantification of uncertainty. For the detection of synchronisation with a large number of potentially relevant time intervals of both signals, ``significant'' effects for specific time point combinations are of particular interest.} Apart from rank based p-values provided in the context of Like\-li\-hood-based boosting \citep{Binder.2009} using permutations of the response, no general inferential framework in the classical statistical sense exists for boosting methods. An alternative approach is stability selection \citep{Meinshausen.2010}, which evaluates the importance of explanatory variables by looking at the stability of term selection under subsampling and has already been adapted for functional regression boosting \citep[see e.g.,][]{Brockhaus.2015}. %and may be very useful if the researcher's interest is, for example, the selection of the most important explanatory variables among a large number of available EEG-signals. %For a detailed description of stability selection in gradient-boosting approaches, see \citet{Hofner.2014}.
In the emotion components application, however, the applied research question defines the chosen covariates and the statistical analysis needs to address the uncertainty of estimated coefficient surfaces. We therefore use a non-parametric curve-level bootstrap to assess the variability of estimated effects. Due to the shrinkage effect of boosting, the corresponding bootstrap intervals are useful for variability quantification of the \textcolor{myc}{regularised} coefficients, but are on average not centred at the true coefficient surface, unlike unbiased estimators. In consequence, the distribution of bootstrap estimates does not provide valid confidence intervals. In the following section, we investigate whether dispite the shrinkage effect, variability bands can be used at least to assess point-wise difference from zero. As simulation results suggest, these variability bands find most of the truly non-zero surface regions in all of our simulation settings.

\section{Simulations} \label{simSection}

%We analyse the performance of the proposed method on the basis of simulated data sets with varying sample sizes, number of trajectories, signal-to-noise ratios, and number of knots for basis functions. 
We provide results for the estimation performance of simple historical effects (subsection \ref{simSubsection1}), %, including settings in which the true effect is multi-modal or restricted to a certain time window. On the one hand, this has the purpose to mimic the data given by the motivating application. On the other hand, we want to examine the performance in settings, where the effect is similar to those imposed by time series methods. 
factor-specific historical effects (subsection \ref{simSubsection2}) and for the uncertainty quantification via bootstrap (subsection \ref{simSubsection3}). In section \ref{sec:furtherSims}, we briefly address results on different parametrisations and boosting step-lengths.%compare estimation for step-lengths $\nu$ in the boosting algorithm being $0.1$ and $1$.

Similar to our application, we use historical effects with integration limits $l(t)=T_1=0$ and $u(t)=t- \delta$ with $\delta = 0.025$. %, a so called \emph{lead effect}, relating all previous time points but not the last $\delta$ time units to the current value of $Y$. 
%Estimation performance and bootstrap based uncertainty quantification were evaluated for up to five different main historical effects and two factor-specific historical effect surfaces. 
%Random factor-specific and random historical effect surfaces were generated by drawing normal random coefficients for tensor product B-splines (see subsection \ref{simSubsection2} for more details). 
We compare the estimated surface with the underlying true function and, wherever possible, with an estimate using a functional additive mixed model as implemented in the \texttt{pffr}-function in R package refund \citep{Scheipl.2015}. Apart from visual comparisons, we estimate the relative integrated mean squared error (\emph{reliMSE}) $\iint ( \hat{\beta}(s,t) - \beta(s,t) )^2 \, ds\,dt  \cdot (\iint \beta(s,t)^2 \,ds\,dt)^{-1}$ by its discrete approximation in order to compare the estimates of our method, referred to as \texttt{FDboost}.

Simulation settings were generally based on $n \in \lbrace 80, 160, 320, 640 \rbrace$ number of observed curves with $D_i \equiv D \in \lbrace 20, 40, 60 \rbrace$ observed grid points per trajectory, a \emph{signal-to-noise ratio} $\text{SNR} \in \lbrace 0.1, 1, 10 \rbrace$. For the following subsections the combinations were customised or restricted accordingly, in particular for simulations with very time-consuming bootstrap calculations. Whereas the number of curves in our application $n=184$ is within the range of simulated settings, we use fewer observations per trajectory in our simulations than available in our application ($D = 384$) in order to reduce computational time. Increasing \textcolor{myc}{sampling density} $D$ from $60$ to $180$ or $380$ in additional simulations with $\text{SNR} \in \lbrace 0.01, 0.1, 1 \rbrace$ and $n = 160$ almost always results in an improvement of estimation performance. The average estimated SNR in our application is $0.42$, which, due to the shrinkage effect, might potentially be underestimate the true SNR. \textcolor{myc}{We also present results of another simulation for $n \in \lbrace 24, 48 \rbrace$, $D \in \lbrace 190,380 \rbrace$ and $\text{SNR} \in \lbrace 0.01, 0.1, 1 \rbrace$ in the appendix. The results suggest that, even for a small number of observations, the estimation performance is satisfactory when the density of sampling is large enough.} %  with an absolute reduction of up to about $0.2$ for the median reliMSE.

The results of our simulation studies are briefly summarised in the following sections. See the supplementary material for a full presentation of results.

\subsection{Estimation of historical effects} \label{simSubsection1}

\textcolor{myc}{Though estimation performance for simple historical effects has already been examined in \citet{Brockhaus.2016}, we provide additional simulation results for complex multi-modal effect surfaces. The simulation settings are motivated by} our application, in which several time windows may show a relationship between the two biosignals. We thus simulate data sets where the effect surface is multi-modal for both the $s$- and the $t$-direction. %To produce realistic functional observations, we also examine the estimation performance when adding an oscillating measurement error as present in the application particularily for the EMG-signal due to voltage fluctuations. 
%For different data generating mechanisms and settings, we compare the logarithmic reliMSE of \texttt{FDboost}- and \texttt{pffr}-fits.
Samples were generated from the model
\begin{equation}
Y_i(t) = \alpha(t) + \int_0^{t-\delta} x_i (s) \beta (s,t) ds + \varepsilon_i(t), \quad i=1,\ldots,n, \label{dgp1}
\end{equation}
for which the functional covariate $x_i(s)$ is simulated as sum of $\varkappa \in \lbrace 5,7,9,11 \rbrace$ natural cubic B-Splines with independent random coefficients from a standard normal distribution. The true underlying coefficient surface is given by $\beta(s,t) = \sin( 10 \cdot |s-t|) \cdot \cos(10t) I(s \leq t-\delta)$ with $I(s \leq t-\delta) = 1$ if $s \leq t-\delta$, else $0$. The independent Gaussian error process $\varepsilon(t)$ with mean zero has constant variance $\sigma^2$ defined via the $\text{SNR} =  \sqrt{\textnormal{Var}(\Xi)} / \sqrt{(\sigma^2)}$ with $\textnormal{Var}(\Xi)$ being the empirical variance of the linear predictor. %In addition to the model error, a high frequency oscillating measurement error $\tau(t) = \sin(1500\,t)/20$ is added to the linear predictor in order to imitate the specificities of the measurement instruments.

In addition, we simulate effect surfaces with a band structure. This is done by using the data generating process in (\ref{dgp1}) and restricting the influence of $x_i$ to values $s$, for which $s \leq t - \delta$, $s \geq t - 0.1$ and $t \leq 0.75, s,t \in [0,1]$. With $40$ observed time points the restriction $s \geq t - 0.1$ corresponds to an autoregressive model with time-varying effects and a lag of $0.1/(1/40) = 4$ time points. With this simulation, we want to investigate whether our approach is able to adequately recover the effect of $x_i$ restricted to a certain number of lags without having to predefine lags. This would be an advantage over time-series models which have to specify the assumed lag structure a priori and would allow a corresponding dimension reduction without restricting the analysis. 

\textit{Results}. \textcolor{myc}{For combinations in which $n$ and $\text{SNR}$ are not very small at the same time,} our gradient-boosting approach works well and recovers the true underlying functional relationship. These findings are depicted in Figure \ref{fig:bandPlot}. %the following figures for $n = 640$ and $\text{SNR} = 1$. 
As can be seen in the upper row, both \texttt{pffr} and \texttt{FDboost} are able to recover the true underlying effect well (right panel) with \texttt{FDboost} having an advantage for low $\text{SNR}$ and low $n$ (left panel). For higher $\text{SNR}$, where \texttt{FDboost} shows less of an improvement than \texttt{pffr} compared to the low $\text{SNR}$ setting, boosting estimates may potentially be further improved by using a higher number of iterations (limited to $1500$ for this subsection). %For this subsection the maximum number of stopping iterations was limited to $1500$ for computational reasons and chosen via $15$-fold bootstrap resampling. For very low $\text{SNR}$ values, a large number of iterations may lead to overfitting, but choosing the optimal number of stopping iterations seems to give good protection against this. 
In the supplementary material, we additionally provide estimates with average reliMSE for a smaller number of observations, visualising the deterioration in estimation performance with decreasing sample size. %In order to obtain good results, the number of knots has to be chosen appropriately high to prevent oversmoothing. Additionally, estimates of both methods showed better performances for marginal first than for marginal second difference penalties of the B-spline basis.

Similar to the multi-modal example, \texttt{FDboost} outperforms \texttt{pffr} (lower row of Figure \ref{fig:bandPlot}) for band surfaces in settings with a lower SNR, whereas for a $\text{SNR} = 10$, \texttt{pffr} shows partly better performances. As exemplarily shown in the lower right panels of Figure \ref{fig:bandPlot}, \texttt{FDboost} is often able to correctly detect the non-zero regions, whereas the typical estimated surface of \texttt{pffr} exhibits larger parts with false positive estimates. 
\begin{knitrout}
\definecolor{shadecolor}{rgb}{0.969, 0.969, 0.969}\color{fgcolor}\begin{figure}[ht]

{\centering \includegraphics[width=\maxwidth]{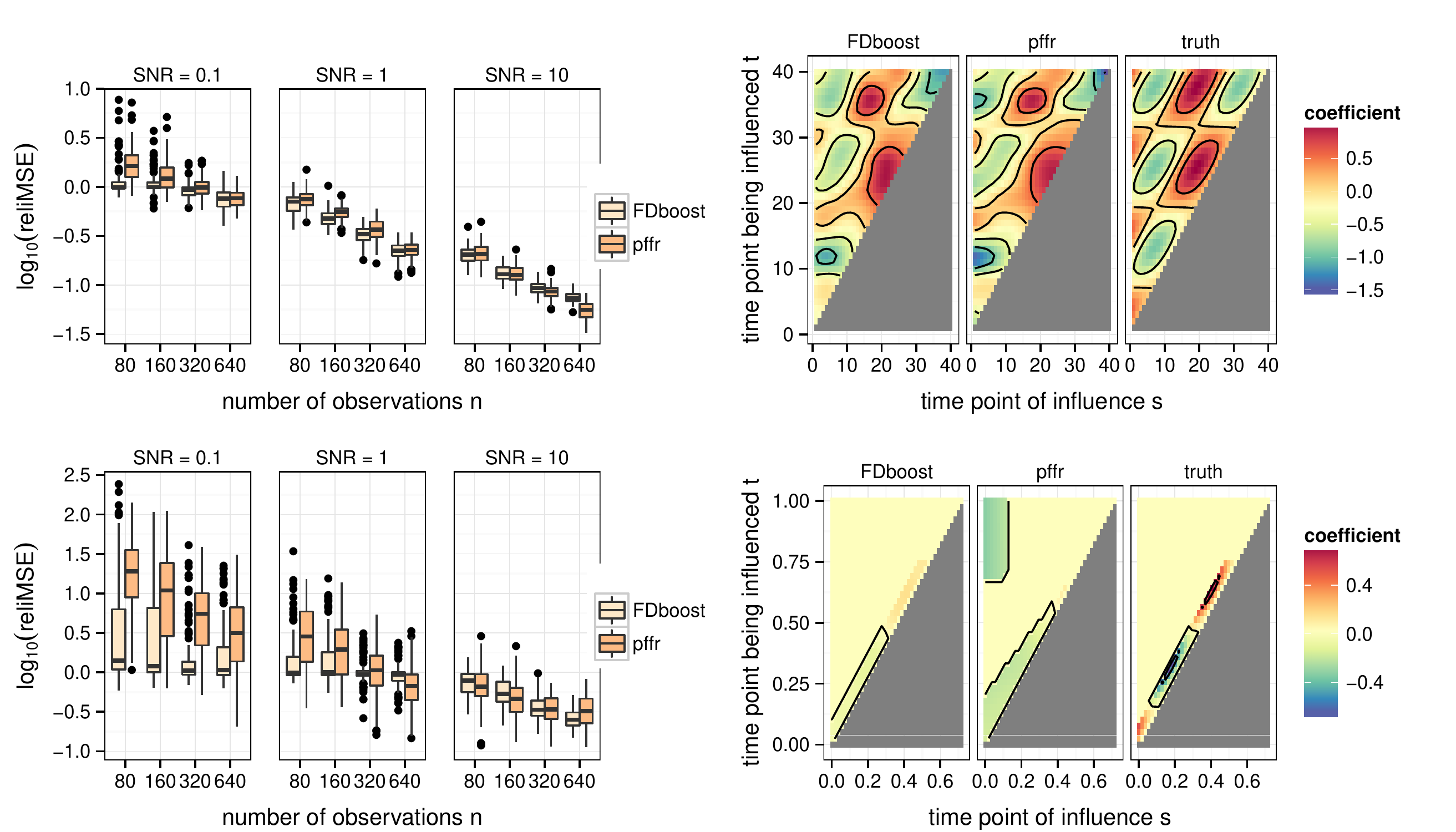} 

}

\caption{Left panels: Comparison of reliMSEs for the estimation of multi-modal surfaces (upper row) as well as band surfaces (lower row) and different settings of the SNR (columns). The $x_i(s)$ were generated on the basis of $11$ natural cubic B-Splines with $15$ knots. Right panels: example for estimates of a multi-modal surface (upper row) for $640$ observed trajectories and a SNR of $1$ respectively estimates of a band structured coefficient surface (lower row) for $320$ observed trajectories and a SNR of $10$ both with respective average reliMSE.}\label{fig:bandPlot}
\end{figure}

\end{knitrout}
%
%\noindent In general, boosting yields excellent results when the majority of relevant effects is not located at the edges of the surface. However, if strong relationships are present for very small or large values of $s$ and $t$, i.e. at the beginning or end of the observed signals, the performance of boosting can sharply deteriote. We come back to this problem in the course of subsection \ref{simSubsection3}.

%\subsection{Estimation of interaction effects} 
\subsection{Estimation performance for factor-specific historical effects} \label{simSubsection2}

%The main focus of this section lies in the estimation performance for factor-specific or random historical effects. %Therefore the simulation of historical effects, which vary with a certain covariate, was done as follows. 
For random historical effects, we adapt the ideas of \citet{Scheipl.2016} and \citet[][Web Appendix C]{Brockhaus.2016} and generate random coefficient functions $\beta_{f}(s,t)$ as linear combinations of cubic P-splines \citep{Eilers.1996} for $n_{\text{subject}} = 10$ factor levels (subjects). % with spline coefficients drawn from a normal distribution and with covariance corresponding to first or second order difference penalties in both $s$- and $t$-direction. 
The coefficient functions $\beta_{f}(s,t), f = 1, \ldots, \varphi = 10$ are then centred to comply with constraint (\ref{sumToZeroConstr}). For factor-specific historical effects, we specify multiples $\iota(e)$ of one fixed coefficient function $\varpi(s,t) = \frac{s}{\sqrt{2}} \cdot \cos(\pi \sqrt{t})$ \textcolor{myc}{with $\iota(e)$ being centred coefficients drawn uniformly between $-5$ and $5$ for each factor level $e = 1, \ldots, \eta = 4$}, allowing for a more systematic examination of estimation accuracy in specific regions of the coefficient function. An additional doubly-varying effect is simulated by multiplying $\varpi(s,t)$ with centred random coefficients drawn from a standard normal distribution. 

In a first series of settings (correctly specified case), the data are generated on the basis of the fitted model, including a main historical effect and (i) a time-varying categorical effect as well as a factor-specific historical effect, (ii) a time-varying random effect as well as a random historical effect, (iii) combining (i) and (ii), or (iv) combining (iii) with a doubly-varying historical effect (full model). In a second series of settings, the model is misspecified by fitting a single historical effect, whereas the data are simulated using a main and (v) a factor-specific historical effect or (vi) a random historical effect or alternatively (vii) by generating the data from the full model whereas the model is fitted without the doubly-varying effect.
% 
% 
% 
% \begin{enumerate}[(i)]
% \item a time-varying categorical effect as well as a factor-specific historical effect, \label{item:a}
% \item a time-varying random effect as well as a random historical effect, \label{item:b}
% \item (\ref{item:a}) as well as (\ref{item:b}) \label{item:c} or
% \item (\ref{item:c}) as well as a doubly-varying historical effect (full model). \label{item:d}
% \end{enumerate}
% and simulation scenarios, in which the model is misspecified by fitting a single historical effect, whereas the data is simulated using a main and 
% 
% \begin{enumerate}[(i)]
% \setcounter{enumi}{4}
% \item a factor-specific historical effect or 
% \item a random historical effect, 
% \end{enumerate}
% as well as 
% 
% \begin{enumerate}[(i)]
% \setcounter{enumi}{6}
% \item generating the data from the full model whereas the model is fitted without the doubly-varying effect.
% \end{enumerate}
% 

\textit{Results}. %Figure \ref{fig:iaFullPlot} exemplarily shows the performances when data is generated on the basis of the full model (iv). For factor-specific historical effects there are no available methods other than \texttt{FDboost} for comparison. 
%
%
%As for the main historical effect model from the previous subsection, boosting reveals very good performances for the main and factor-specific historical effect.  %In general, the performance for random factor-specific historical effect estimation seems to be dependent on the actual factor-specific historical effect. Based on a full model fit including a main historical effect as well as both single-varying and the doubly-varying historical effect, Figure \ref{fig:iaFullPlot} shows the comparison of estimation performances if the model is correctly specified. 
Whereas the main historical effect for the settings (i)-(iv) shows a similar logarithmic reliMSE as in previous simulation settings in \ref{simSubsection1}, the historical effects varying with a categorical covariate show more diverse performances and larger deviations. The factor-specific and random historical effect estimation mostly capture the main features of the true underlying surface, but are not estimated as reliably as the main historical effect. Estimates for the doubly-varying historical effect %Due to the large number of factor levels ($\eta \cdot \phi = 40$) the coefficients of the random factor-specific historical effect are often 
are often shrunk almost to zero due to an insufficient number of observations. 
In settings (v) or (vi) where the true underlying model includes a random or factor-specific historical effect, estimation performance for the main historical effect is equally good when fitting the correct or the misspecified model. For setting (vii) the performance is practically the same for the estimation of the main historical effect. The difference in estimation performance varies more strongly for the factor-specific as well as random historical effect and, in particular, indicates a better performance of the correctly specified model for high $\text{SNR}$ and larger $n$. The fact that estimation performance is not affected more strongly is likely due to the orthogonality of the omitted effect to the effects included in the model, cf. (\ref{sumToZeroConstr}) - (\ref{sumToZeroConstr3}).

\subsection{Quantification of uncertainty} \label{simSubsection3}

%We further investigate the properties of bootstrap based uncertainty statements for historical effects. %Due to the shrinkage effect of boosting, the empirical bootstrap distribution is biased towards zero and uncertainty intervals, based on this distribution, while capturing the variability of the estimates, share this bias. For this same reason, the coverage of bootstrap intervals does not preserve the nominal level. %In the case of a high signal-to-noise ratio and very stable estimations within each bootstrap sample, intervals may even have a coverage of zero percent due to the shrinkage effect. 
In the following, we examine the ability of $95\%$-bootstrap intervals to correctly identify (non-)zero coefficients in the manner of conventional confidence intervals by looking at the inclusion of zero. On the basis of $100$ nonparametric bootstrap iterations, we calculate the %\emph{sensitivity}, the \emph{specificity}, 
\emph{false negative rate} (\emph{FNR}) and \emph{false positive rate} (\emph{FPR}) over the surface for each of $100$ simulated data sets. In addition, the frequencies of false negative (\emph{FFN}) and false positive estimates (\emph{FFP}) for each surface point across all data sets are obtained. \textcolor{myc}{We present results for a model including only one main historical effect in addition to a model with main and factor-specific historical effects}, for both of which true coefficient surfaces are partly equal to zero. The true coefficient surface for the main historical effect is defined as $\beta(s,t) = \mathcal{Q}_{0.001}\lbrace\sin(|t-s|+10) \cdot \cos(5s)\rbrace$ and surfaces for factor-specific historical effects are simulated as multiples of $\varpi(s,t) = \mathcal{Q}_{0.001}\lbrace\phi_{0.9,0.2}(s) \cdot \phi_{0.9,0.2}(t)\rbrace$, where $\mathcal{Q}_a(x) = x \cdot I(x \geq a)$ and $\phi_{\mu,\sigma}(\cdot)$ is the normal density function with expectation $\mu$ and variance $\sigma^2$. We additionally investigate the performance of our uncertainty quantification for a model including main and random historical effects, which are simulated as described in section \ref{simSubsection2}.

\textit{Results}. Figure \ref{fig:boxplotFalseRates1} depicts the results for a simple historical effect simulation with $\text{SNR} = 1$, $n=160$ and $D=40$. Both the FNR and the FPR are below $0.05$ in all but a few cases. When decreasing the $\text{SNR}$ to $0.1$, the bootstrap approach yields smaller FPR at the cost of a larger FNR. 
Considering the FFP and FFN, %which are calculated for each single coefficient surface point over all simulation iterations, 
$8$\% of all non-zero surface points reveal a FFN of above $0.05$ and $30$\% of all zero surface points reveal a FFP of above $0.05$. Plotting the FFN against the coefficient size indicates that FFNs larger than $0.05$ only occur for coefficient values of below $0.2$ (below 0.6 if $\text{SNR} = 0.1$). The rightmost panel of Figure \ref{fig:boxplotFalseRates1} reveals a strong relationship between the FFP and a smaller distance to non-zero points on the surface, with FFP mostly below about $0.1$ for points not next to a non-zero coefficient. 
\begin{knitrout}
\definecolor{shadecolor}{rgb}{0.969, 0.969, 0.969}\color{fgcolor}\begin{figure}[ht]

{\centering \includegraphics[width=\maxwidth]{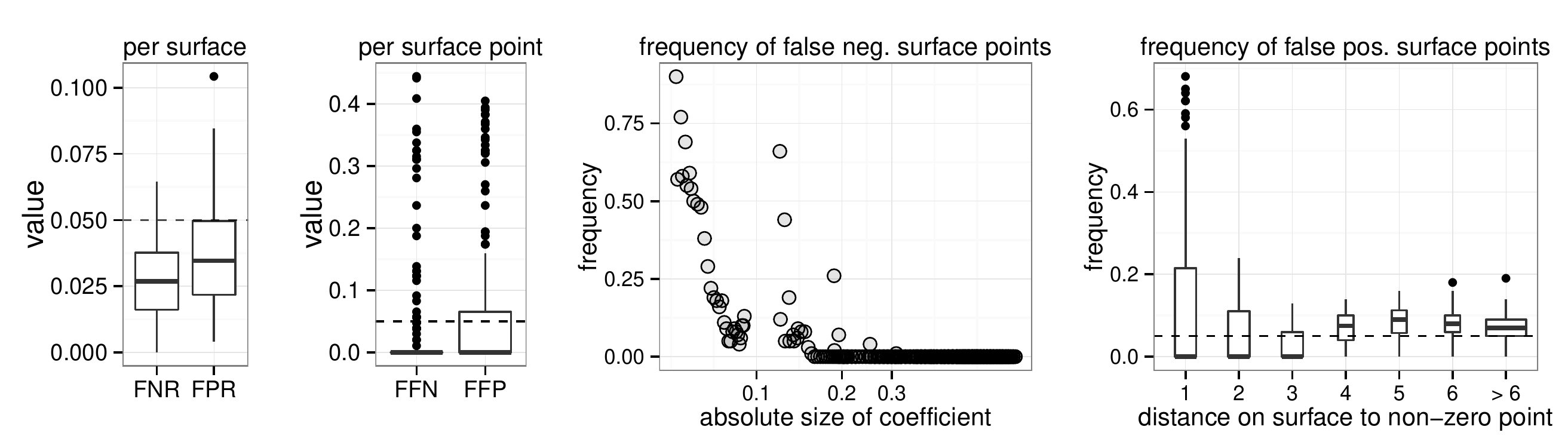} 

}

\caption{Results for uncertainty quantification of a simple historical effect and data generated with $S\!N\!R=1$, $n=160$ and $D=40$. FNR and FPR for each surface with boxplot over iterations (left panel) as well as FFN and FFP for each surface point over all simulation iterations with boxplot over surface points (second panel). Third panel: frequency of bootstrap intervals including zero plotted against the coefficient size for each truly non-zero coefficient surface point. Forth panel: frequency of false positive estimates plotted against the minimal distance to a true non-zero point for each zero surface point.}\label{fig:boxplotFalseRates1}
\end{figure}

\end{knitrout}
Though the performance depends on the specific surface, the bootstrap approach finds the majority of non-zero coefficient points in simulations for a simple historical model and tends to have a FFN of almost zero. A large FFP only occurs for surface points, that are directly adjacent to true non-zero coefficient points.

For a more complex model also including a factor-specific historical effect, the bootstrap approach works well regarding the detection of the truly non-zero surface area. However, it reveals considerably higher FNR as well as higher FFN particularly for smaller coefficients of both effect surfaces.
\textcolor{myc}{In the case of correlated observations, for example given by repeated measurements per subject, we subsample on the level of independent observation units (subjects)}. In the simulation with a main and a random historical effect, higher frequencies of false positive estimates for the main historical effect occur, which, however, are again located around the true non-zero coefficient area. %Increasing the number of subjects from $10$ to $20$ in an additional simulation study hardly affects this fact. indicating that the deterioration of uncertainty quantifaction performance is mostly due to the small number of independent observation units.  %requires a sufficient number of independent observation units to obtain reliable results.

In summary, simulation results suggest that the bootstrap approach does not comply with the chosen confidence level in the manner of conventional confidence intervals, but proves to find most of the truly non-zero surface regions for all simulation settings. Large FFN and FFP are mainly revealed at the edges of non-zero coefficient areas, such that an interpretation of detected non-zero areas of the surface are still possible as long as exact pixel locations of edges are not taken at face value.

\subsection{Further simulations} \label{sec:furtherSims}

In addition to the presented simulations, we investigate the performance of boosting for different parameterisations as introduced in section \ref{sec:facSpecHist} and compare boosting estimates with step-length $\nu = 0.1$ and $\nu = 1$. The gradient boosting algorithm is defined for step length $\nu \in (0,1]$. In general, it is recommended to set the step length ``sufficiently small'' \citep{Buehlmann.2007} for predictive accuracy reasons, for example in the range of $0.01$ and $0.1$. A larger step length and, in particular $\nu = 1$, requires much fewer iteration steps and therefore speeds up the model fit, but may result in a deterioration of prediction performance due to overfitting. Since we are rather interested in the estimation performance of model components, we investigate whether or how much overfitting is a problem in our particular setting.

\textit{Results}. For the two different parameterisations, performances differ on a relatively small scale, suggesting that the choice of parameterisation can be based on the given research question. In the comparison of step-lengths, there appears to be no clear best choice in all settings. Thus estimation with $\nu = 1$ might be a reasonable alternative to smaller step-lengths, requiring less computing time and memory consumption due to a smaller number of necessary iterations, especially in complex models applied to large data sets.

\section{Application to the detection of synchronisation in bioelectrical signals} \label{application}

\subsection{Data and background} \label{sec5.1}

%Synchronisation of psychophysiological signals plays an important role in understanding the elicitation of emotion episodes. 
%Addressing the question of the temporal dynamics of emotion eliciting appraisal processes, 
\citet{Gentsch.2014} conducted a study in which $24$ participants played a computerised gambling game with real monetary outcome. During the gambling rounds, \citeauthor{Gentsch.2014} modified three factors (so-called appraisals) related to Scherer's Component Process Model \citep[CPM, ][]{Scherer.2009} and simultaneously recorded brain activity with EEG and facial muscle activity with EMG. %The EMG-measurements represent (1) the frontalis muscle, which raises the eyebrows, (2) the corrugator muscle, which causes a frown and (3) muscles of the cheek region associated with smiling or activation of mouth corners. 
In componential emotion theories such as the CPM, an emotion episode is assumed to emerge through the synchronisation of the emotion components (e.g., appraisals, expressions, or feelings). 
%which are highly interrelated as the results of the appraisal of a given situation or event trigger changes in each response component. When these become highly synchronised an emotion arises. 
In order to investigate synchronisation processes, % between these components, 
\citeauthor{Gentsch.2014} operationalised three dichotomous appraisals, which are included as dummy variables in the present data set: (1) \emph{Goal conduciveness}, which was related to the monetary outcome at the end of each gambling round (\textit{gain} coded as $G=1$ or \textit{loss} with $G=0$), (2) \emph{Power}, which allowed players to change the final outcome if the setting was \emph{high power} (\emph{hp} for short coded as $P=1$, else referred to as \emph{low power} / \emph{lp} with $P=0$) and (3) \emph{Control}. \textcolor{myc}{The control setting was manipulated in blocks in order to change the participant's subjective feeling about her ability to cope with the situation}. Before a block with several gambling rounds would start, participants were told whether they were going to have high or low power for the majority of upcoming games, which corresponds to \emph{high} or \emph{low control} settings (\emph{hc} coded as $C=1$ respectively \emph{lc} with $C=0$). In rounds with high control, for example, the player was told to frequently have high power\textcolor{myc}{, thereby trying to induce a subjective feeling of control over the situation,} and vice versa for low control. \textcolor{myc}{Each participant played over $100$ gambling rounds for each of the eight appraisal settings, which we also refer to as \emph{trials}.} %Thus, the data set at hand includes the three dummy variables \textit{goal conduciveness} ($D = 1$ if the round was won, else $0$), \textit{power} ($P = 1$ if the power was high, else $0$) and \textit{control} ($C = 1$ if the control was high, else $0$) possibly interacting with each other (which yields eight different settings) and 
%To illustrate these data, Figure \ref{fig:explData} exemplarily shows the EEG- (upper row) and the EMG-signal for one participant, averaged over all trials for each of the eight possible game conditions. It is evident that the (averaged) EMG-signal (representing the frontalis muscle activity, which raises the eye brows) is relatively noisy, whereas the (averaged) EEG-signal (representing the fronto-central electrode \emph{Fz}, in particular measuring intentional and motivational activities, \citet{Teplan.2002}) reveals a smoother function and shows more variation for different settings. 

Before performing statistical analyses, EEG- as well as EMG-signals are pre-pro\-cessed (see the supplementary material for further details). After removing the data of one participant due to considerably deviating observations, \textcolor{myc}{which imply a defective or displaced sensor}, several hundred gambling rounds each with $384$ equally spaced EEG- and EMG-measurements within around 1500 milliseconds are available for each of the 23 participants. 
\textcolor{myc}{Analogous to previous studies on synchronisation and, in particular, the study of \citet{Gentsch.2014}, we use aggregated observations for each participant and game condition by averaging the corresponding trials for each time point. On the one hand, this results in less computing time and the feasibility to quantify uncertainty in effect estimates via bootstrap, on the other hand, this is motivated by investigations on event-related potentials (ERPs). ERP analysis is a commonly practised method to infer from neuronal activity. Neuronal activity is thought to be time-locked in delay to a certain stimulus, wherefore aggregating over a large number of trials is used to cancel out random brain activity and strengthens those parts of the signal, which are commonly observed for all trials \citep[see, e.g.,][]{Pfurtscheller.1999, Handy.2005, Rousselet.2008}.}
%Analogous to previous studies on synchronisation, we use aggregated observations for each participant and game condition by averaging the corresponding trials for each time point. This is in line with other studies investigating event-related potentials (ERPs) \citep[see, e.g.,][for more details on ERP analysis]{Pfurtscheller.1999}.

\textcolor{myc}{Instead of combining the (spatially correlated) EEG-signals in order to maximize the explanatory power of the analysis, the question of interest rather lies in the dominant influence of certain selected EEG-signals. We fit a model for each EEG-signal of interest ($Fz$-, $FCz$-, $POz$- and $Pz$-electrode) in order to determine the direct effect on the facial muscle activity. In order to demonstrate the capability of our approach to handle high-dimensional data sets, we also provide sample code in the repository for fitting a model, in which all $64$ EEG-signals are potentially included with historical, factor-specific and random historical effects. In the supplementary material, we additionally provide a visualisation for the selection frequency of this model after $2000$ iterations.}

\subsection{Model}

It is predicted that facial expression is largely driven by efferent brain signals reflecting appraisal processes. We use the following maximal model
\begin{equation}
Y_{il} (t) = \sum_{j=1}^{13} h_j(\bm{x}_{il})(t) + \varepsilon_{il}(t), \label{eq:modelAppl1}
\end{equation}
for $l=1,\ldots, n_{\text{setting}} = 8$, $i=1,\ldots,n_{\text{subject}} = 23$, $t \in \mathcal{T} = [0ms,1500ms]$ and $D_i \equiv D = 384$ observed time points in $\mathcal{T}$. In (\ref{eq:modelAppl1}), $Y_{il}(t)$ represents a chosen EMG-signal for subject $i$, game condition $l$ and time point $t$ in the game. $h_j(\bm{x}_{il})(t)$, or, for short, $h_j(t)$ are thirteen partial effects of covariates $\bm{x}_{il}$ including a time-varying intercept, game condition effects ($C$, $P$, $G$) and EEG-signal effects depending on the selected electrode signal $\omega_{il}$. Table \ref{tab:tab01} provides the details on each part of the linear predictor. For the integration limits, we use $l(t)=0$ and a lead-parameter $u(t) = t - \delta = t - 12ms$, which is meaningful due to restrictions given by the neuro-anatomy of humans and is just below the time lag between EMG and EEG of $14.3ms$ \citep{Mima.1999}. In order to reflect subject-specific variation, we include time-varying random intercepts and subject-specific historical EEG-effects in the model.

\begin{table}[ht]
\caption{\label{tab:tab01}Partial effects in the EMG-EEG-model}
\centering
\fbox{%
\begin{footnotesize}
\begin{tabular}{ll}
\emph{Partial effect} $h_j(x_{il})(t)$ & \emph{Effect (of)} \\ \hline
$h_1(t) = \alpha(t)$ & Intercept \\
$h_2(t) = b_{0,i}(t)$ & Subject-specific intercepts \\
\, & \,  \\
$h_3(t) = \gamma_1(t)C_{il}$ & Game condition $control$ \\
$h_4(t) = \gamma_2(t)P_{il}$ & Game condition $power$ \\
$h_5(t) = \gamma_3(t)G_{il}$ & Game condition $goal$ $conduciveness$ \\
$h_6(t) = \gamma_4(t)C_{il} P_{il}$ & Interaction of $control$ and $power$ \\
$h_7(t) = \gamma_5(t)C_{il} G_{il}$ & Interaction of $control$ and $goal$ $cond.$ \\
$h_8(t) = \gamma_6(t)P_{il} G_{il}$ & Interaction of $power$ and $goal$ $cond.$ \\
$h_9(t) = \gamma_7(t)C_{il} P_{il} G_{il}$ & Interaction of all game conditions \\
\, & \,  \\
$h_{10}(t) = \int_{0}^{t-12} \omega_{il}(s) \beta_1(s,t)\, ds$ & EEG-signal \\
$h_{11}(t) = \int_{0}^{t-12} \omega_{il}(s)\, \beta_{2,l}(s,t)\, ds$ & EEG-signal (game-condition specific) \\
$h_{12}(t) = \int_{0}^{t-12} \omega_{il}(s)\, b_{1,i}(s,t)\, ds$ & EEG-signal (subject-specific)\\
$h_{13}(t) = \int_{0}^{t-12} \omega_{il}(s)\, b_{2,i,l}(s,t)\, ds$ & EEG-signal (subj.- and game cond.-spec.) \\
\end{tabular}
\end{footnotesize}
}
\end{table}

% The effects listed in Table \ref{tab01} are particularly suitable for answering the given research question as in the CPM subjective feelings are assumed to emerge after a critical threshold in synchronisation of appraisal-driven changes \citep{Scherer.1984}. This theoretical behaviour is directly translated to the model at a given time point $t$ by integrating the (individually weighted) EEG-signal impacts for all preceding time points $s < t - 12$. 

%To reflect subject-specific variation, we include time-varying random intercepts and subject-specific historical EEG-effects in the model. %In addition to the individual reaction to the given game conditions, voltage values of the EEG data may differ across participants due to confounding variables \citep{Cohen.2014}. %Subject-specific values, for example, might be caused by different skull shapes or cortical folding \citep{Cohen.2014}, which are difficult to explicitly account for or even to collect. 

Though game condition-specific historical effects may well be subject specific, simulations in the previous section suggest that even if the true model corresponds to the full model, estimation performance is only slightly affected when using a misspecified model without a random factor-specific historical effect $h_{13}(t)$. %Moreover, by aggregating the data per subject and game condition, it is hardly possible to estimate the partial effect $h_{13}(t)$. 
As a sensitivity analysis, we also fit the full model including $h_{13}(t)$ on a finer aggregation of the data, for which we average over fewer trials per subject and thus obtain repeated measurements per subject-game condition-combination.%, yielding $5$ or $10$ observed curves for each combination of participant and game condition. The coarsest aggregation (one curve per subject and game condition) results in $n = n_{\text{subject}}  \cdot n_{\text{setting}} = 23 \cdot 8 = 184$ observed curves and $D = 384$ grid points per trajectory. 
% The coarsest aggregation is richer than most of the simulations for $D$ while $n$ lies close to the simulated setting with $n = 160$ observations. 

\subsection{Results}

For the historical effects, the estimated coefficient surfaces are depicted in Figure \ref{fig:resultPlots1a} for the EEG-covariate in the form of the electrode 'Fz' (in particular measuring intentional and motivational activities, \citet{Teplan.2002}) and the EMG-response signal of the frontalis muscle (raises the eyebrows). The lower panel in these figures depicts the average EEG-signal per game condition, demeaned per time point by the overall mean and with negative or positive values highlighted in blue or red, respectively. Two further panels (left, center) for the EMG-signal show the overall mean, the prediction with and without the historical effects (left) as well as the difference between these predictions (center). For predictions, the average EEG-signal per game condition was used. Additionally, corresponding bootstrap results for uncertainty assessment are incorporated in the figures by different degrees of transparency related to different pointwise bootstrap intervals $BI_\alpha = [q_{\alpha/2},q_{1-\alpha/2}]$, $q_a$ as $\alpha$\%-bootstrap quantile and $\alpha \in \lbrace 1,5,10 \rbrace$. Surface points are coloured with the \textcolor{myc}{corresponding} coefficient value and are less transparent if the specified bootstrap interval does not contain the value zero.% For less transparent points, corresponding coefficient values therefore are more consistently non-zero in the bootstrap replicates. 
Figure \ref{fig:resultPlots1a} shows the sum of the estimated coefficient surfaces of main and game condition-specific historical effects for the four high control settings (the other four surfaces are included in the online appendix). In all four effect surfaces a similar pattern can be found, which reflects the structure of the main historical effect. The coefficients near the diagonal reveal a positive sign at around $s \approx 500ms$, whereas the upper left as well as the upper right of the surface, visually separated by a thick black contour line, are estimated with a negative sign. In contrast to the upper left negative coefficient area, which is mostly indicated to be not different from zero by the boostrap, the upper right negative coefficient area is indicated to be non-zero for all eight conditions at least to some extent. The positive area in between those two negative subareas is mostly estimated to be either zero or non-zero but with relatively small coefficient values. The positive effect near the diagonal at $s \approx 500ms$ is estimated to have the largest values for hc settings in combination with hp / loss and lp / gain situations and is found to be non-zero by the bootstrap only for the latter scenario. This very strong short-term synchronisation of EEG- and EMG-signal seems to be very reasonable from a theoretical point of view, as facial reactions including raising of the eyebrows are usually of brief nature and are linked to appraisals such as novelty, which is consistent in the hc / lp / gain case with low power not being expected in a high control setting \citep{Scherer.2009}.

The estimated effect can on the one hand be interpreted on the subject level. A person with a higher EEG-signal at $s \approx 500ms$, for example, will on average show a higher EMG-signal (i.e. stronger muscle activity) for $t \approx 600ms$, given the preceding EEG-signal and game condition remain the same. On the other hand, effects can be \textcolor{myc}{explained} by relating the demeaned average EEG-signal for one game condition and the corresponding coefficients to the changes in the average EMG-signal, which is illustrated by the hc / lp / gain setting in Figure \ref{fig:resultPlots1a}. As EEG values related to this game condition are on average above the overall mean EEG values for $s\in [300,1000]ms$, the EEG seems to have an increasing effect on subsequent EMG values and thus muscle activity, with the effect lasting for at least $100ms$. %Additionally, the influence of a specific time point $s$ in the EEG-signal on the EMG-signal extends for larger values of $t$ up to $s+400ms$. 
%
% From $s \approx 500ms$ and $t \approx 600ms$ on the positive effect of the EEG is decreasing and instantaneous effects change the sign for $s \approx 900ms$ or larger. From a theoretical point of view, however, these effects are rather naturally given correlations than an actual existing synchronisation. 
%
%Additionally, another small part of the coefficient surface with negative signs for $s \in [1200,1300]ms$ was found to be non-zero. The game condition-specific average of EEG values being smaller than the overall mean in this interval, this indicates a slightly increasing but almost instantaneous effect of late EEG values on the EMG values. 
% For the EMG-signal, the model predicts an increase in EMG values which is mainly independent of this particular EEG-signal (difference of grey and green line in the prediction plot), but is also partly related to this EEG-signal (difference of green and black line). Both contributions are additionally plotted in the difference plot, where the positive effect from around $s = 350ms$ to $s = 1000ms$ is particularly visible. 

In theory, muscle activity should be traceable to brain signals. Therefore the results indicate that brain activity measured at the $Fz$-electrode only contributes to a relatively small amount in explaining the movement of eyebrows (difference panels on the left of each plot in Figure \ref{fig:resultPlots1a}). % and that muscle activity%, which is not traced back to the EEG-signal (but a certain game condition), is due to signals of other brain regions not covered by the $Fz$-electrode. 
However, for the game condition hc / lp / gain, the model explains a considerable amount of EMG-activity (particularly visible in the difference plot of EMG predictions). 

When reparameterising the factor-specific historical effects without historical main effect, when boosting with step-length $1$ as well as in the full model with more finely aggregated data, the estimated effects are similar to the reported ones. Further results for the application are given in the online appendix, including results for the scalar covariates. 

\citet{Gentsch.2014} analysed EEG- and EMG-signals separately and made statements regarding differences in game conditions for one of the signals at a time. \textcolor{myc}{Although this and other similar strategies may yield results on significant changes in one signal for different study settings, no statement on the association of the two signals can be made.} In contrast, investigating the emotion components data with our proposed approach facilitates the modeling of synchronisation of EMG- and EEG-signals in the first place and additionally allows the simultaneous EEG- and EMG-analysis to differ for influence factors given by the study design. \textcolor{myc}{Our method therefore is able to recreate parts of the theoretical emotion components model and leads to new insights on the underlying synchronisation process. Specifically, we found associations between EEG- and EMG-signal that are time-localized (without the need to prespecify time lags) and which differ between experimental settings, with setting hc / lp / gain showing the clearest association.}

\begin{knitrout}
\definecolor{shadecolor}{rgb}{0.969, 0.969, 0.969}\color{fgcolor}\begin{figure}[ht]

{\centering \includegraphics[width=\maxwidth]{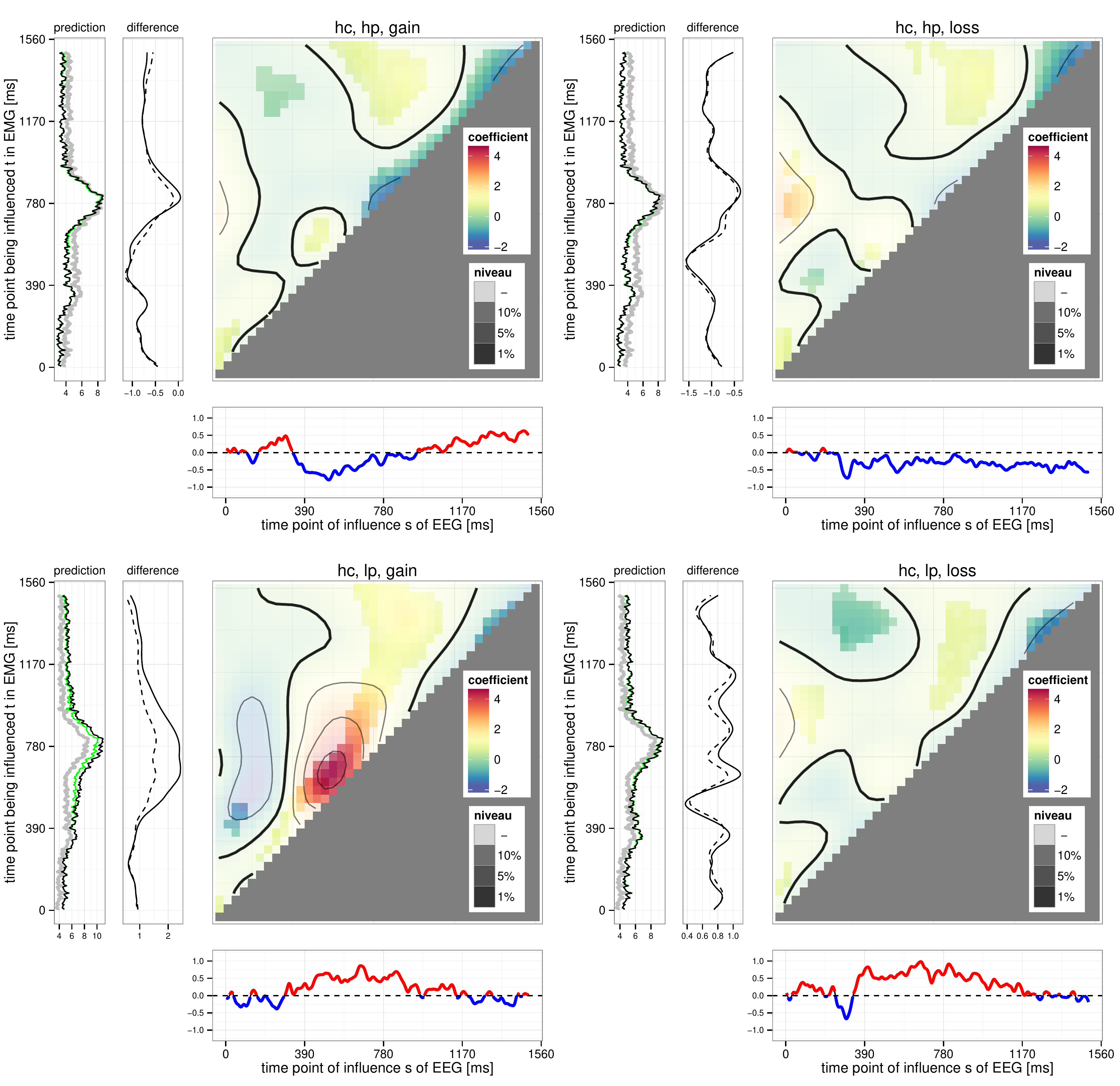} 

}

\caption{Estimated coefficient surfaces for the model with EEG-covariate 'Fz' (plot of average signals per game condition at bottom with negative and positive values highlighted \textcolor{myc}{blue and red, respectively}; signals are demeaned per time point by the overall mean), all four high control settings and the EMG-response signal of the frontalis-muscle (left panels: overall mean (1) in grey, prediction without historical effects (2) in green, with historical effects (3) using the average EEG-signal per game condition in black; center panel: dashed line as difference between (1) and (2), solid line as difference between (1) and (3)). Surfaces correspond to estimated main historical effect plus game condition specific historical effect. Different degrees of transparency in coefficient plot indicate surface points having $(1-$niveau$)$-bootstrap intervals which do not contain the value zero. \textcolor{myc}{To obtain a reasonably sized image estimated effects are visualised on a $40 \times 40$ grid}}\label{fig:resultPlots1a}
\end{figure}

\end{knitrout}

\section{Discussion}

The focus of this paper is the development of a regression framework for the synchronisation analysis of bioelectrical signal data. Bioelectrical signals like EEG or EMG are recorded in many different research areas, as for example, in neuroscience or cognitive neuropsychology, where the goal is to develop an understanding of synchronisation processes in emotion episodes. \textcolor{myc}{In contrast to previous approaches, which are mostly based on coherence, cross-correlation or similar concepts \citep[see, e.g.,][]{Mima.1999, Brown.2000, Grosse.2002}, we use a function-on-function regression model \citep[see, e.g.,][]{Morris.2015} with factor-specific historical effects}. Our model extends the simple historical model \citep{Malfait.2003, Harezlak.2007, Brockhaus.2016} by factor-specific and / or random historical effects. As far as we know, there are no methods available other than \texttt{FDboost} allowing historical effects to vary with other covariates. We develop constraints to make the resulting estimates both interpretable as well as identifiable. This flexible class of function-on-function regression models is implemented in the R package \texttt{FDboost}. Using the component-wise gradient boosting \textcolor{myc}{approach by \citet{Brockhaus.2015, Brockhaus.2016}} for estimation, this approach can deal with high-dimensional data, even $p > n$ settings, and includes variable selection. %\textcolor{myc}{This is not possible in other implementations such as the \texttt{pffr}-function in the R package \texttt{refund}.} %or the \texttt{linmod}-function in the R package \texttt{fda}. 
\textcolor{myc}{The algorithm} is able to recover different effect surfaces, including relationships assumed in time series approaches, and allows for potentially time-varying associations. The quality of estimates is comparable to those of the function \texttt{pffr} of the R package \texttt{refund} for special cases of function-on-function regression where \texttt{pffr} is applicable.

%Simulations show good estimation performance for various settings. Whereas the direction of the effect surface is correctly detected for reasonable sample sizes and signal-to-noise ratios, boosted model coefficients exhibit a shrinkage effect, such that absolute coefficient sizes have to be interpreted with care. 
A bootstrap can be employed to assess the variability of boosted estimates. While bootstrap intervals, due to the shrinkage, do not constitute confidence intervals with proper coverage, simulations show that the bootstrap approach is able to recover areas with non-zero effects very well and only shows a larger FPR and FNR at the edges of true non-zero effect surfaces. A better uncertainty quantification would be a relevant avenue for future developments. %The major challenge in this context is a statement on the coefficient size of parts of the coefficient surface simultaneously taking the shrinkage effect of boosting into account. 

%We additionally investigated the behaviour of boosted historical models for different parametrisation and step-lengths in the boosting algorithm. As it turns out, estimation performances may be improved by using no penalisation for the factor variable part in factor-specific historical effects (if possible) and only specifying a factor-specific historical effect without a main historical effect. For models with full step-length it seems, that the performance is only affected if the signal-to-noise ratio of the given data is relatively small. Therefore, computation time and memory usage may be drastically reduced by setting the step-length to one, thereby potentially obtaining almost identical results as, for example, demonstrated with our application.\\ %Model selection criteria do not seem to be very promising with respect to further reduce computation time or memory usage due to expensive matrix multiplications. An exception may be given in settings with a modest number of observations for which cross-validation proves difficult. \color{blue} 

While we do not focus on this feature here, our approach can also model other characteristics of the conditional response distribution than the mean, such as the median or a quantile. A more complex yet interesting class of models would be obtained by combining functional regression models with generalized additive models for location, scale and shape as done for scalar response by \citet{Brockhaus.2016b}.% This would provide a way not only to model the conditional mean of the response functions, but also their variance or other features of the conditional response distribution. %Another extension may be given by incorporating the warped functional regression approach of \citet{Gervini.2014}, for example explicitly allowing typical features of game effects in our application to be shifted for each observed trajectory, thereby however forfeiting the interpretability of effects with respect to fixed time periods.\\
% \section*{Supplementary Materials}

For the emotion components data, our model contributes to the understanding of the componential theory by estimating a functional relationship between the EEG and EMG signals without having to prespecify a certain time lag between these two signals. In addition, our proposed extension for historical models allows for appraisal-specific investigations on synchronisation processes of emotion components.

% \vspace*{0.5cm}
% 
% \noindent \textbf{Acknowledgements}
% 
% \noindent We thank Fabian Scheipl for his help and useful comments. Sonja Greven, Sarah Brockhaus and David R\"ugamer acknowledge funding by Emmy Noether grant GR 3793/1-1 from the German Research Foundation. Kornelia Gentsch and Klaus Scherer were funded by an ERC Advanced Grant in the European Community’s 7th Framework Programme under grant agreement No. 230331-PROPEREMO to Klaus Scherer and by the National Center of Competence in Research (NCCR) Affective Sciences financed by the Swiss National Science Foundation (No. 51NF40-104897) hosted by the University of Geneva.

\section*{Acknowledgements}
We thank Fabian Scheipl for his help and useful comments. Sonja Greven, Sarah Brockhaus and David R\"ugamer acknowledge funding by Emmy Noether grant GR 3793/1-1 from the German Research Foundation. Kornelia Gentsch and Klaus Scherer were funded by an ERC Advanced Grant in the European Community’s 7th Framework Programme under grant agreement No. 230331-PROPEREMO to Klaus Scherer and by the National Center of Competence in Research (NCCR) Affective Sciences financed by the Swiss National Science Foundation (No. 51NF40-104897) hosted by the University of Geneva.

\bibliography{mybibliography}

\clearpage

\end{document}